\documentstyle[pra,floats,aps,epsf]{revtex}

\begin{document}
\draft
\title{Dynamical TAP approach to mean field glassy systems}
\author{Giulio Biroli}
\address{Laboratoire de Physique Th{\'e}orique de l'Ecole Normale Sup{\'e}rieure
\thanks{Unit{\'e} Mixte de Recherche du Centre Nationale de la Recherche Scientifique et de l'Ecole Normale Sup{\'e}rieure.},\\
24 rue Lhomond, 75231 Paris cedex 05, France
}
\maketitle

\begin{abstract}
The Thouless, Anderson, Palmer (TAP) approach to thermodynamics of mean field
 spin-glasses is generalised to dynamics. A method to compute the dynamical
 TAP equations is developed and applied to the p-spin spherical model.
 In this context we show to what extent the dynamics 
 can be represented as an evolution in the {\em free energy} landscape.
 In particular the relationship
 between the long-time dynamics and the local properties
 of the {\em free energy} landscape shows up explicitly within 
 this approach. Conversely, by an 
 instantaneous normal modes analysis we show that the local  
 properties of the {\em energy} landscape seen by the system 
 during its dynamical evolution do not change qualitatively at
 the dynamical transition.

\end{abstract}

\pacs{PACS Numbers~:75.10.Nr-64.70.Pf-61.43-j. Preprint LPT-ENS 99/23}

%\narrowtext

%%%%%%%%%%%%%%%%%%%%%%%%%%%%%%%%%%%%%%%%%%%%%%%%%%%%%%%%%%%%%%%%%%%%%
%  Intro
%%%%%%%%%%%%%%%%%%%%%%%%%%%%%%%%%%%%%%%%%%%%%%%%%%%%%%%%%%%%%%%%%%%%%
 If at large times a system relaxes toward
 the equilibrium state, its dynamics is called ``equilibrium dynamics'' and 
 equilibrium properties as the fluctuation dissipation
 relation and the time translation invariance hold. 
In this case the departures of 
 the dynamical probability measure from the Gibbs measure vanish at 
large times, therefore the relationships between thermodynamics 
 and long-time dynamics are obvious. \\
However there are many physical cases, in which a system remains far from
 equilibrium also at long times. In the following we focus
 on glassy systems, for which relaxation times become so long 
 at low temperature that these systems are not in equilibrium on laboratory
 time scales \cite{rivista}. In this case it is important to understand
 to what extent pure static concepts (e.g. the free energy landscape) can be 
 related to the long-time dynamics. \\
For thermodynamics the {\em relevant} landscape 
 is the free energy one. Different authors \cite{landscape}
 proposed 
 that this landscape is {\em relevant} also for dynamics and 
 can be considered, at least at the
 simplest level, as the landscape on which the dynamical evolution
 takes place. For example for a spin 
system the landscape
 for dynamics would be the free energy as a function of local magnetisations
 $m_{i}(t)$ and the dynamical variable would be the set of the averaged
 (over different thermal histories) local magnetisations . At the
 simplest level the dynamical evolution would be a 
superposition of two 
 phenomena: a gradient descent in this free energy landscape and jumps 
between different states with
 a probability given by a generalised Arrhenius law: $\exp(-\beta 
\Delta F) $, where $\Delta F$ is the free energy barrier between 
 two states. \\
However this is far from obvious and up to now a proof of these 
 claims is not available. The main difficulty is that in general
 the free energy landscape is not known and the long-time dynamics
 is not solved.\\
This ``Landscape Paradigm'' \cite{landscape} has received a firm 
theoretical basis in the case of mean field frustrated systems, 
 for which an analytic solution 
of the thermodynamics \cite{beyond}
 and of the asymptotic out of equilibrium regime \cite{rivista} is
 in general available. \\
For these models it was shown that a complicated 
 energy function can lead to a rugged free energy  
 landscape and to an infinite number of correlated states. 
Thouless, Anderson and Palmer (TAP) \cite{TAP} computed for the Sherrington Kirkpatrick model \cite{SK}
 the free energy as a function of local magnetisations. At low temperature
 the TAP free energy 
 has an infinite number of minima; each one corresponds to a different possible
 state. It has been shown that their 
weighted sum (with the Boltzmann weight)  gives back equilibrium
 results\cite{young}. Moreover states are correlated and organised
 in a ultra-metric structure. This is encoded in 
the Parisi's solution \cite{Parisi} and was 
explicitly shown in the cavity approach 
\cite{cavita} developed by M{\'e}zard, Parisi and Virasoro,
 which is a way to solve the minimisation equations of the TAP free energy.\\
Furthermore at low temperature 
these systems remain out of equilibrium also at very large
 times \cite{rivista} and their long-time dynamical behaviour
 exhibits non trivial features like
 violation of the fluctuation-dissipation theorem
 and ageing \cite{leticia}. \\
In particular for the p-spin 
spherical model Cugliandolo and Kurchan \cite{leticia} showed 
 that, even if the system remains always out of equilibrium, the long-time
 dynamical behaviour can be interpreted 
 in terms of some properties of the free energy landscape.
 The most intriguing fact is that the properties of the free energy landscape
 relevant for
 long time dynamics and thermodynamics are completely different. These  
 results indicate that, at least in this mean field case, 
 there is a close relationship between long-time dynamics
 and the free energy landscape, which therefore has a meaning on its own
 also in an out of equilibrium regime. 
Besides, we note that a connection between
 the free energy landscape and the long-time out of equilibrium dynamics
 is very interesting not only for its theoretical implications, but also
 from a technical point of view. In fact this relationship allows
 to obtain results about dynamics
 by a pure static computation \cite{Remi,Franz}.\\
However, the reason of this relationship is not clear. 
Is the description of the dynamics
 as an evolution on the free energy landscape correct or does something
 else happens, but always 
such that the relationship found in \cite{leticia} between
 the asymptotic behaviour and the free energy landscape
 is satisfied ?\\
Up to now an answer to this question has been given only for the 
zero temperature Langevin dynamics of p-spin ($p\geq 2$) spherical 
models \cite{laloux}. In this case there is
 no thermal disorder, so it is clear that a landscape over which 
 the dynamics takes place exists and is the energy landscape
 (or the free energy one, because
 at zero temperature they coincide). In \cite{laloux} it
 was shown that at
 least at zero temperature the main reason of ageing is the
 flatness of the landscape seen during the long-time dynamics . \\
The regularity of the dynamical
 equations near $T=0$ and the interpretation of the asymptotic behaviour
 in terms of TAP free energy \cite{leticia} seem to indicate that the above
 scenario might be true also at non zero
 temperature. However, in this description of ageing 
 it is implicitly assumed that at large times 
 a landscape for dynamics should exist and 
that this landscape should be related
 to TAP free energy. Thus, the question
 about the relationship among dynamical behaviour and free energy
 landscape arises again.

 In this article we clarify this relationship for the
 p-spin spherical model. The thermodynamical and the dynamical
 behaviours of this model exhibit strong analogies
 with the phenomenology of super-cooled liquids, the glass
 transition and the glassy phase \cite{vetri1,rivista,vetri2}. 
Moreover the dynamical theory of the p-spin spherical model
 has a close relationship \cite{mcep-spin} 
with the Mode Coupling theory \cite{gotze},
 which serves as a basis for some theories of super-cooled
 liquids.
 For these reasons many authors consider that the p-spin spherical model
 provides a mean field description of the glass transition and
 of the glassy phase.

 For this model the TAP free energy was computed and studied
 in detail \cite{ptap,crisantitap} and an analytic solution of 
 the asymptotic out of equilibrium regime is available 
\cite{leticia,crisantidyn}. To understand the reasons of the connection
 found in \cite{leticia} between TAP free energy and the asymptotic
 behaviour, we will compute the equations satisfied by the local magnetisations
 $m_{i}(t)=\left<s_{i}(t)\right>$ (where $\left< \cdot \right>$ 
means the average over
 the thermal noises) without performing the average over 
 disorder. This is the generalisation to dynamics of 
the Thouless, Anderson and   
Palmer approach \cite{TAP}.\\ 
We will show that the dynamical evolution of the local magnetisations
 corresponds to a relaxation 
 in the free energy landscape (in a sense which we will
precise) only for very large times and for particular initial conditions;
 in all the other cases the dynamics is characterised by a memory term, which 
 makes the evolution non-Markovian. Moreover the study of the  dynamical
 TAP equations shows that the stationary points of
 the static free energy and the
 free energy Hessian in these points are closely related to the 
 long-time dynamical behaviour, as was already found from
 the solution of the equations on the correlation and the response
 functions in \cite{leticia,Franz,Alain}. Our results explicitly show
 that the scenario for slow dynamics found in \cite{laloux} remains valid
 also at finite temperature: ageing is due to the motion in the flat
 directions of the free energy landscape in presence of a vanishing source
 of drift.\\
Finally we show that already for the simple case of the p-spin spherical 
model an analysis of the local properties of the {\em energy} 
landscape is not adequate to
 identify the dynamical glass transition. We will compute the spectrum
 of the energy Hessian for dynamical configurations seen
 during the dynamical evolution. 
The eigenvectors
 of the energy Hessian are called instantaneous normal modes in liquid
 theory and have been introduced to represent the short-time dynamics
 of liquids within a harmonic description \cite{keyes}. \\ 
We will show that the local properties
 of the energy landscape seen during the dynamical evolution
 do not change qualitatively at the dynamical glass transition
 but at a higher temperature $T_0$, which seems to be related
 to the damage spreading 
transition \cite{ritort}.
This indicates that 
 at the dynamical glass transition the energy landscape
 seen by the system remains locally the same, whereas its {\em global}
 properties
 change and this can be observed by analysing the {\em local} properties 
of the free energy landscape.
 
The paper is organised as follows: in section I we derive the static
 TAP free energy for mean field spin glass models. 
This section serves as an introduction to the method
 applied in the following to derive the dynamical TAP free energy. 
 In section II this method is applied to derive the dynamical TAP
 equations via the analogy between the dynamical theory and a 
 supersymmetric static theory. In section III the asymptotic analysis 
 of the dynamical TAP equations is performed. In section IV the
 local properties of the 
energy landscape seen during the dynamical evolution
 is analysed. Finally we conclude in
 section V.
%%%%%%%%%%%%%%%%%%%%%%%%%%%%%%%%%%%%%%%%%%%%%%%%%%%%%%%%%%%%%%%%%%%%%
%  tap statiche
%%%%%%%%%%%%%%%%%%%%%%%%%%%%%%%%%%%%%%%%%%%%%%%%%%%%%%%%%%%%%%%%%%%%%
\section{Static TAP approach}

A useful function in the study of phase transitions 
 is the Legendre transform of free energy.
 This function, which can be interpreted as the effective potential whose
 minima represent different possible states, 
gives an intuitive (and quantitative) description of phase transitions. 
 Consider for example ferromagnetic systems. 
In this case the effective potential
 is a function of magnetisation. The ferromagnetic transition 
 corresponds to the splitting of the paramagnetic minimum in 
 the two ferromagnetic minima. A vanishing external magnetic field breaks
 the up-down symmetry and fixes the system in one of the two possible
 ferromagnetic states.\\
Generally, frustrated systems are characterised by a complicated energy
 landscape, which can give rise eventually to the existence of
 many possible states. In this case it is not possible to characterise
 the states a priori studying the various possible schemes of spontaneous
 symmetry breaking. Thus, the effective potential has to be computed 
as a function of the averaged microscopic
 configuration and this is what Thouless, Anderson and Palmer did
 for the Sherrington-Kirkpatrick model. They derived the Legendre transform 
 of the free energy with respect to the 
local magnetic fields $h_{i}$, obtaining
 the effective potential (called now TAP free energy) which is a function
 of the local magnetisations $m_{i}$ for a fixed (but typical) 
 disorder configuration. At low temperature the TAP free energy 
 has an infinite number of minima; each one corresponds to a different possible
 state. It has been shown that their 
weighted sum (with the Boltzmann weight)  gives back equilibrium
 results\cite{young} found by the replica \cite{Parisi} or the cavity method\cite{beyond}. 

In this section we show 
how the static TAP free energy can be derived for mean field spin
glass models. In this way we present in a simple case the strategy 
which we will follow to compute dynamical TAP
 equations. The derivations of the static TAP equations presented 
 in the literature \cite{tapstatiche,crisantitap} seem to be quite often model dependent
 and rather involved. 
Therefore we hope that this section may be also useful to
 show an easy way to obtain the TAP free energy for
 a generic mean field model. It does not matter if one considers
 spherical or Ising spins, two body or many body interactions.

A straightforward way to
 compute the TAP free energy for mean field (completely connected)
 spin glass models consists in finding a good perturbative
 expansion such that using the properties of typical disorder
 configurations it is possible to show that 
 only a finite number of terms of the perturbation
 series does not vanish. \\
Following this strategy we will compute the Legendre transform of the free
 energy with respect to $\left<S_{i}\right>$ and 
$\sum_{i=1}^{N}\left<S_{i}^{2}\right>$. One
 may wonder why we Legendre 
 transform also with respect to $\sum_{i=1}^{N}\left<S_{i}^{2}\right>$;
 the reason is that otherwise the
 perturbation expansion would contain an infinite number of terms.\\
Note that for Ising spins the 
average $\sum_{i=1}^{N}\left<S_{i}^{2}\right>$ is trivial,
 so we will Legendre transform only with respect to magnetisations.
 This is a peculiarity of the Ising spins, which disappears
 for spherical (or Potts) spins and in the dynamical case.

\subsection{Static TAP equations}\label{tapstat}
We are finally in position to define the TAP free energy 
$\Gamma (\beta , m_{i},l)$ which depends on the 
magnetisation $m_{i}$ at every site $i$ and on the spherical
 parameter $l$ (for Ising spins $l$ is absent). 
$\Gamma$ is the Legendre transform of the
``true'' free energy:    
%There are many different ways to obtain this result 
%\cite{tapstatiche}, we 
%will follow the simple and physical approach developed by T. Plefka 
%\cite{Plefka} and A. Georges and S. Yedidia \cite{Antoine} for the 
%Sherrington-Kirkpatrick model.
\begin{equation}\label{gdef}
-\beta \Gamma (\beta , m_{i},l)=\ln 
%\underset{{S_{i}}}{Tr} 
 \mbox{Tr}_{\{S_{i}\}} \exp \left(
-\beta H(\{S_{i}\}) - \sum_{i} h_{i} (S_{i}-m_{i}) -\frac{\lambda}{2} 
\sum_{i=1}^{N} (S_{i}^2-l)\right)\quad .
\end{equation} 
For Ising spins $\mbox{Tr}_{\{S_{i}\}}
=\sum_{\{S_{i}\}}$ and for spherical spins
$\mbox{Tr}_{\{S_{i}\}}=\int_{-\infty}^{+\infty} \prod_{i=1}^{N} dS_{i} $.
The Lagrange multipliers $h_{i}(\beta)$ fix the magnetisation at each site $i$:
 $\left<S_{i}\right>=m_{i}$ and $\lambda(\beta)$, which is present only for spherical
models, enforces the condition
$\sum_{i=1}^{N} \left<S_{i}^2-l\right>=0$. $\left< \cdot \right>$ denotes the thermal average 
  and $N$ is the number 
of spins. \\
Once $\Gamma $ is known, the equation $-\frac{2}{N} 
\frac{\partial \beta \Gamma}{\partial l} {\Big \vert} _{l=1}=
\lambda$ fixes the spherical constraint ($\sum_{i}S_{i}^{2}=N$) and 
gives the spherical multiplier as a function of $m_{i}$, whereas 
$-\frac{\partial \beta \Gamma}{\partial m_{i}} {\Big \vert} _{l=1}=
h_{i}$ are the TAP equations, which fix the values of local magnetisations.\\
In the following we focus on the p-spin Hamiltonian:
\begin{equation}\label{hp-spin}
H(\{S_{i}\})=-\sum_{1\leq i_{1}<\cdots <i_{p}\leq N} J_{i_{1},\cdots,i_{p}}
S_{i_{1}}\cdots S_{i_{p}}\quad ,
\end{equation} 
where the couplings are Gaussian variables with zero mean and average
 $\overline{J_{i_{1},\cdots,i_{p}}^{2}}=\frac{p!}
{2N^{p-1}}$.

The standard perturbation expansion for the generalised potential $\Gamma $
 is rather involved \cite{doubletransform,crisantitap} and cannot be
 directly applied to the Ising case. Thus, we prefer to follow 
the approach developed for the 
Sherrington-Kirkpatrick model by T. Plefka 
\cite{Plefka} and A. Georges and S. Yedidia \cite{Antoine} 
because is simple and can be
directly applied to all mean field spin glass models. \\
They obtained the TAP free energy for the Sherrington-Kirkpatrick model 
expanding
 $-\beta \Gamma$ in powers of
 $\beta$ around $\beta =0$:
\begin{equation}\label{expgamma}
-\beta \Gamma (\beta , m_{i},l)=\sum_{n=0}^{+\infty}-\frac{\partial ^{n} 
(\beta \Gamma)}{\partial \beta ^{n}}{\Bigg \vert}_{\beta = 0}
\frac{\beta ^{n}}{n!}\quad .
\end{equation} 
For a general system this corresponds to a 
$\frac{1}{d}$ expansion ($d$ being the spatial dimension) around mean
 field theory \cite{Antoine}; so it is not surprising that for mean 
field spin glass models only a finite number of terms survives. The 
zeroth- and first-order terms give the ``na{\"\i}ve'' TAP free energy, whereas
 the second term is the Onsager reaction term.

From the definition of $-\beta \Gamma$ given in equation (\ref{gdef}), 
we find for Ising spins:
\begin{equation}\label{g0ising}
-\beta \Gamma _{I}(\beta, m_{i}) {\Bigg \vert}_{\beta =0}=-\sum_{i}^{N}
\left[ \frac{1+m_{i}}{2}\ln\left(\frac{1+m_{i}}{2} \right)+\frac{1-m_{i}}{2}
\ln\left(\frac{1-m_{i}}{2} \right) \right]
\end{equation}
and for spherical spins\footnote{We are neglecting a useless constant in
 $\Gamma _{S}$. A term in $\Gamma _{S}$, that does not depend 
on $l$ and $m_{i}$, has no influence on thermodynamics.}:
\begin{equation}\label{g0sferi}
-\beta \Gamma _{S}(\beta, m_{i},l) {\Bigg \vert}_{\beta =0}=\frac{N}{2}
\ln \left( l-\frac{1}{N}\sum_{i=1}^{N} m_{i}^2 \right)\quad .
\end{equation}
These are the entropies of non interacting Ising or 
spherical spins constrained to have magnetisation $m_{i}$.\\
The linear term in the power expansion (\ref{expgamma}) of the
 TAP free energy equals:
\begin{eqnarray}\label{g1a}
-\beta \frac{\partial (\beta\Gamma _{S,I})}{\partial \beta} {\Bigg \vert}
_{\beta =0}&=&\beta \sum_{1\leq i_{1}<\cdots <i_{p}\leq N} J_{i_{1}
,\cdots,i_{p}} \left<S_{i_{1}}\cdots S_{i_{p}}\right>_{\beta=0}-\\ 
&&-\beta \sum_{i}^{N} 
\frac{\partial h_{i}}{\partial \beta}{\Bigg \vert}
_{\beta =0} \left<S_{i}-m_{i}\right>_{\beta=0}
-\beta  
\frac{\partial \lambda}{\partial \beta}{\Bigg \vert}
_{\beta =0} \sum_{i=1}^N \left<S_{i}^2-l\right>_{\beta=0}\quad ,\nonumber
\end{eqnarray}
where the last sum is present only for spherical models.
The second and the third term are zero because of Lagrange conditions;
 moreover at $\beta =0$ 
the spins are decoupled so all the thermal averages are trivial:
\begin{equation}\label{g1b}
-\beta \frac{\partial (\beta\Gamma _{S,I})}{\partial \beta} {\Bigg \vert}
_{\beta =0}=\beta \sum_{1\leq i_{1}<\cdots <i_{p}\leq N} J_{i_{1}
,\cdots,i_{p}} m_{i_{1}}\cdots m_{i_{p}} \quad.
\end{equation}
This ``mean field'' energy together with the zeroth-order term
 gives the standard mean field theory, which becomes exact for infinite-ranged
ferromagnetic system. The Onsager reaction term comes from the quadratic term 
in the expansion (\ref{expgamma}):
\begin{eqnarray}\label{g2a}
 -\frac{\beta ^{2}}{2} \frac{\partial ^{2}(\beta\Gamma _{S,I})}{\partial 
\beta ^{2}} {\Bigg \vert}
_{\beta =0}  &=&  \frac{\beta ^{2}}{2}\left< \left(\sum_{1\leq i_{1}<\cdots
<i_{p}\leq N} Y_{i_{1},\cdots,i_{p}}\right)^{2}\right>^{c}_{\beta =0}\quad ,\\
\nonumber \\
Y_{i_{1},\cdots,i_{p}} &=&  J_{i_{1},\cdots,i_{p}}S_{i_{1}}\cdots S_{i_{p}}-
(S_{i_{1}}-m_{i_{1}})m_{i_{2}}\cdots m_{i_{p}}- \\
&&\cdots -m_{i_{1}}\cdots 
m_{i_{p-1}}(S_{i_{p}}-m_{i_{p}})\quad .\nonumber
\end{eqnarray}
To compute this term we have used the following Maxwell relations:
\begin{eqnarray}\label{maxwell}
\frac{\partial h_{i}}{\partial \beta}{\Bigg \vert}_{\beta =0} & = & 
-\frac{\partial}
{\partial m_{i}}\frac{\partial (\beta\Gamma _{S,I})}{\partial \beta}
{\Bigg \vert}_{\beta =0}\\
\frac{\partial \lambda}{\partial \beta}{\Bigg \vert}_{\beta =0} & = & 
-\frac{2}{N}\frac{\partial}
{\partial l}\frac{\partial (\beta\Gamma _{S})}{\partial \beta}
{\Bigg \vert}_{\beta =0}
\end{eqnarray}
Using the statistical properties of the couplings it is easy to check
 that the only terms giving a contribution of the order of $N$ correspond
 to the squares of $J_{i_{1},\cdots,i_{p}}$:
\begin{eqnarray}\label{g2b}
 -\frac{\beta ^{2}}{2} \frac{\partial ^{2}(\beta\Gamma _{S,I})}{\partial 
\beta ^{2}} {\Bigg \vert}
_{\beta =0}  &=&  \frac{\beta ^{2}}{2}\sum_{1\leq i_{1}<\cdots <i_{p}\leq N} 
\left<Y^{2}_{i_{1},\cdots,i_{p}}\right>^{c}_{\beta =0}\\
&=&\frac{\beta ^{2}}{2 p!} \sum_{i_{1}\neq \cdots \neq i_{p}} J^{2}_{i_{1},\cdots,i_{p}}{\Big (}
\left<S^{2}_{i_{1}}\right>_{\beta =0}\cdots \left<S^{2}_{i_{p}}\right>_{\beta =0}-m^{2}_{i_{1}}
\cdots m^{2}_{i_{p}}\nonumber\\
&& -p\left<(S_{i_{1}}-m_{i_{1}})^{2}\right>_{\beta =0}m^{2}_{i_{2}}\cdots m^{2}_{i_{p}}
{\Big )}\quad .\nonumber
\end{eqnarray}
Using again the statistical properties of the couplings and neglecting
 terms giving a contribution of an order smaller than $N$ we find that the 
reaction term depends on $m_{i}$ through the overlap $q=\frac{1}{N}\sum_{i}m_{i}^{2}$ only:
\begin{eqnarray}\label{g2c1}
 -\frac{\beta ^{2}}{4} \frac{\partial ^{2}(\beta\Gamma _{I})}{\partial 
\beta ^{2}} {\Bigg \vert}
_{\beta =0} & =& \frac{\beta ^{2} N}{2}\left(1-q^p-p(q^{p-1}-q^{p})\right)
   \quad ,\\ \nonumber \\
\label{g2c}
 -\frac{\beta ^{2}}{2} \frac{\partial ^{2}(\beta\Gamma _{S})}{\partial 
\beta ^{2}} {\Bigg \vert}
_{\beta =0} & = & \frac{\beta ^{2} N}{4}\left(l^p-q^p-p(lq^{p-1}-q^{p})
\right)\quad .
\end{eqnarray}
Higher derivatives in equation (\ref{expgamma}) lead to terms which can
 be neglected because they are not of order of N \cite{crisantitap,Antoine};
 so collecting
 (\ref{g0ising}),(\ref{g0sferi}),(\ref{g1b}),(\ref{g2c1})
 and (\ref{g2c})
 we find the TAP free energy for Ising and spherical p-spin models. 
Deriving the free energy with respect to magnetisations $m_{i}$ and the
 spherical parameter $l$ (in the spherical case) one finds the
 TAP equations. For instance for spherical spins we find: 
\begin{equation}\label{tapstatichem}
\frac{m_{i}}{1-q}=\frac{\beta }{(p-1)!}\sum_{i_{2}\neq \cdots \neq i_{p}
(\neq i)}J_{i,i_{2},\cdots , i_{p}}m_{i_{2}}\cdots m_{i_{p}} 
-\frac{\beta ^{2}}{2}p(p-1)q^{p-2}(1-q)m_{i}\quad , 
\end{equation}
\begin{equation}\label{tapstatichel}
\lambda = \frac{1}{1-q}+\frac{p \beta ^{2}}{2}(1-q^{p-1})\quad .
\end{equation}
These equations admit for certain temperatures an infinite
 number of solutions. This is a fundamental characteristic
 and difficulty of mean field spin glasses.

It has been shown that 
the weighted sum of the local
 minima of the TAP free energy gives back equilibrium
 results found by the replica or the cavity method \cite{young,crisantitap}:
\begin{equation}\label{Z}
Z=\sum_{\alpha }e^{-N \beta  f_{\alpha }}\qquad ,
\end{equation}
where $f_{\alpha }$ is the TAP free energy of a stable
 solution $\{m_{i}^{\alpha } \}$ of TAP equations. 
The different TAP states can be grouped with respect to their free energy;
 then the partition function can be rewritten as:
\begin{equation}\label{Zf}
Z=\int df e^{-N(\beta f-\Sigma (f;\beta ))}\qquad
\end{equation}
where $N\Sigma (f;\beta )$ is the logarithm of 
the number of TAP states with free
 energy $f$ and is called complexity \cite{rem,vetri1,crisantitap}.\\
Note that states which do not have the minimum free energy can dominate
 the sum in (\ref{Zf}) if their number is very large.

Finally, we note that it was crucial to Legendre transform also
 with respect to $\frac{1}{N}\sum_{i=1}^{N}S_{i}^{2}$, otherwise
 in (\ref{expgamma})
 the derivatives higher than the second order
 would not vanish and an infinite number of terms should be re-summed.
 This re-summation is automatically achieved
 if one Legendre transform also with respect to 
 $\frac{1}{N}\sum_{i=1}^{N}S_{i}^{2}$.
%%%%%%%%%%%%%%%%%%%%%%%%%%%%%%%%%%%%%%%%%%%%%%%%%%%%%%%%%%%%%%%%%%%%%%%%%   
\subsection{A brief survey on the p-spin spherical
%%%%%%%%%%%%%%%%%%%%%%%%%%%%%%%%%%%%%%%%%%%%%%%%%%%%%%%%%%%%%%%%%%%%%%%%%
 model}\label{p-spin}
In the following we recall very briefly some results on the thermodynamics
 and the dynamics of the p-spin spherical model
which will be useful for the asymptotic analysis 
 of the dynamical TAP equations. A detailed review has been
 done by Barrat \cite{Alain3}.
\subsubsection{The thermodynamics}\label{thermo}
The thermodynamics of the p-spin spherical model has been studied
 within the replica approach in \cite{crisantirep}, whereas the TAP
 equations have been analysed in \cite{ptap,crisantitap}.\\
It has been shown that there is a
 high temperature regime $T>T_{d}$, 
 $T_{d}=\sqrt{\frac{p (p-2)^{p-1}}{2 (p-1)^{p-1}}}$, for which
 the paramagnetic 
state $m_{i}=0$ dominates the partition function, i.e. 
$Z=e^{-N\beta f_{para}}$. \\
There is also an intermediate temperature
 regime $T_{s}<T<T_{d}$ in which the partition function is dominated 
 by an exponential number (in $N$) of states, i.e. the related 
 complexity is different from zero. In this case the
 free energy is given by:
\begin{equation}\label{comple}
f_{eq}=f^{*}-\frac{1}{\beta }\Sigma (f^{*};\beta )\qquad 
\beta = \frac{\partial \Sigma }{\partial f}(f^{*};\beta )\quad . 
\end{equation}
Finally, there is a low temperature regime $T<T_{s}$ in which the
 sum in (\ref{Zf}) is dominated by the lowest states in free energy. 
Their number is infinite but not exponential in $N$, so the related
 complexity is zero.\\
In the intermediate temperature regime there is a non trivial
 distribution of states, but the equilibrium free energy ($f_{eq}$) 
 is equal to
 the paramagnetic one ($f_{para}$). Note that the paramagnetic state
 does not exist in this temperature regime \cite{Alain}, but the
 equality between $f_{eq}$ and $f_{para}$
implies that the system seems to be 
 in the paramagnetic phase yet. Actually, in the simplest 
 replica analysis this phase
 is still described by the replica symmetric solution. \\
The thermodynamic phase transition is at $T_{s}$. 
At this temperature the one step
 replica symmetry breaking solution becomes stable and for $T<T_{s}$
 the system is in the glassy phase.
\subsubsection{The dynamics}\label{dyna}
The dynamics of the p-spin spherical model for random 
initial conditions (corresponding to a quench from infinite temperature)
 has been studied in \cite{leticia,crisantidyn}; whereas
 the dynamics taking an initial condition, which is a typical equilibrium configuration at a certain temperature $T'$, has been analysed 
in \cite{Franz,Alain}.\\
It has been shown that for random initial conditions, actually
 the physical case, the p-spin spherical model has a transition at
 $T_{d}$ (which is higher than $T_{s}$). 
Above the dynamical transition temperature there is a coexistence of some 
non trivial TAP states with the 
paramagnetic state which dominates thermodynamics. Starting from a random
 initial condition the system thermalizes within the paramagnetic state, 
however an initial condition belonging to a stable TAP state leads always 
to an equilibrium
 dynamics inside this state. Between the static and the dynamic transition
 temperatures the paramagnetic state is fractured into many  TAP states. 
Starting from a random initial condition the system ends up ageing and the
 asymptotic values of some one time quantities are equal to the corresponding
 ones of the threshold states, which are the highest (in free energy) TAP 
 states. For lower temperatures the static is dominated by the lowest TAP
 states, whereas the dynamics is still dominated by the highest TAP
 states. \\
 Moreover
 the peculiarity of the spectrum of the free energy Hessian for 
threshold states, which is a shifted semi-circle law with minimum 
eigenvalue equal to zero, has been used to give an interpretation 
of ageing as the evolution in a landscape with many flat directions 
\cite{leticia,laloux}.
%%%%%%%%%%%%%%%%%%%%%%%%%%%%%%%%%%%%%%%%%%%%%%%%%%%%%%%%%%%%%%%%%%%%%
%  intro SUSY
%%%%%%%%%%%%%%%%%%%%%%%%%%%%%%%%%%%%%%%%%%%%%%%%%%%%%%%%%%%%%%%%%%%%%
\section{Dynamical TAP approach}
The dynamics of the p-spin spherical model will be investigated
using a Langevin relaxation dynamics.
In the following we introduce the superspace formalism. 
Within this compact notation the 
dynamics and the static theory considered in the previous 
section are formally very similar \cite{kurfranz}. Therefore dynamical 
TAP equations 
 can be derived  straightforwardly generalising the method
 described in the previous section. 
 
\subsection{Formalism}
We start by considering a Langevin equation for the relaxation dynamics of 
spin glass models:
\begin{equation}\label{langevin}
\frac{d s_{i}}{dt}=-\beta \frac{\partial H}{\partial s_{i}}+\eta_{i}(t)\quad ,
\end{equation}
where $\eta_{i}(t)$ are Gaussian random variables 
 with zero mean and variance\footnote{Note that time is measured in unit of 
temperature. The real time is obtained by multiplying $t$ by the inverse
 of the temperature: $t_{r}=\beta t$. This implies that the variance of 
thermal noise is equal to 2 for any temperature.} 
$\left<\eta_{i}(t)\eta_{j}(t')\right>=2 \delta _{i,j}
 \delta (t-t')$. Note that now $\left< \cdot \right>$ means the average
 over the thermal noise.\\
Standard field theoretical 
manipulations \cite{zj} lead to the Martin-Siggia-Rose functional 
\cite{msr} for the expectation value of an operator $O(s_{i})$:
\begin{equation}\label{action0}
\left<O(s_{i})\right>=\int\prod_{i=1}^{N}{\cal D}s_{i} {\cal
D}\hat{s}_{i}
 {\cal D}c_{i} {\cal D}\overline{c}_{i}
\exp \left( S(s_{i},\hat{s}_{i},c_{i},\overline{c}_{i}) \right)O(s_{i})\\
\end{equation}
\begin{equation}\label{action}
S=\int_{0}^{+\infty} dt\left( \sum_{i=1}^{N}-\hat{s}_{i}\left(\frac{ds_{i}}{dt}+\beta \frac{\partial H}
{\partial s_{i}}-\hat{s_{i}}\right)+\sum_{i,j=1}^{N}\overline{c}_{i}\left(\frac{\partial}
{\partial t}\delta_{i,j}+
\beta \frac{\partial ^{2}H}{\partial s_{i}\partial s_{j}}\right)c_{j}\right)
\quad ,
\end{equation}        
where $\overline{c}_{i}(t)$ and $c_{i}(t)$ are Grassmann fields (``ghosts'') 
and $s_{i}(t)$ and $\hat{s}_{i}(t)$ are commuting fields, $s_{i}(t)$ 
 real and $\hat{s}_{i}(t)$ purely imaginary. Starting from the expectation 
value of products of the two 
commuting fields one can construct correlation and response functions:
\[
C(t,t')=\frac{1}{N}\sum_{i=1}^{N}\left<s_{i}(t)s_{i}(t')\right>
\qquad 
R(t,t')=\frac{1}{N}
\sum_{i=1}^{N}\frac{\partial\left< s_{i}(t)\right>}{\partial h_{i}(t')}=\frac{1}{N}\sum_{i=1}^{N}\left<s_{i}(t)
\hat{s}_{i}(t')\right>\quad ,
\]
where $h_{i}(t)$ is the magnetic field coupled to the spin $s_{i}$.
In superspace \cite{zj} the action S can be written in a compact form, which
 looks like a static action. To fix the notation in superspace, we introduce
 two anticommuting Grassmann variables $\theta$, $\overline{\theta}$; the 
integrals over these variables are defined as:
\begin{equation}\label{tetaint}
\int d\theta =\int d\overline{\theta} =0 \qquad \int d\theta \, \theta =
\int d\overline{\theta} \, \overline{\theta}  =1 \quad . 
\end{equation}
We also introduce the (commuting) superfield $S_{i}$:
\begin{equation}\label{superfield}
S_{i}(t,\theta,
\overline{\theta})
=s_{i}(t)+\overline{\theta} c_{i}(t) + \overline{c_{i}}(t) \theta + 
\overline{\theta} \theta \hat{s}_{i}(t)\quad . 
\end{equation}
and the notation:
\begin{equation}\label{notation}
D=-2\frac{\partial ^{2}}{\partial \theta \partial \overline{\theta}}
-2\theta \frac{\partial ^{2}}{\partial \theta \partial t} + 
\frac{\partial}{\partial t},
 \qquad (1) = (t_{1},\theta_{1},
\overline{\theta}_{1})\quad . 
\end{equation}
In terms of superfields the Martin-Siggia-Rose functional 
can be written as (see for example \cite{zj,kurfranz}):
\begin{equation}\label{superaction}
Z=\int \prod_{i=1}^{N}{\cal D}S_{i} \exp\left( \int d1 \left[ -\beta H(S_{i}(1))-
\frac{1}{2}\sum_{i=1}^{N}S_{i}(1)
DS_{i}(1)\right] \right)\quad ,
\end{equation}
where $d1=dt d\theta d\overline{\theta}$.\\
The action $S$ is supersymmetric \cite{zj} and the generators 
of this supersymmetry are:
$D'=\frac{\partial}{\partial \overline{\theta}}$ and $\overline{D}'=
\frac{\partial}{\partial \theta}+\overline{\theta}
\frac{\partial}{\partial t}$.
The first generator implies in particular that the action $S$ is invariant 
with respect to translations of $\overline{\theta}$. By using the hypothesis
 that the SUSY is not broken it is possible to show \cite{gozzi} that 
 the system is at equilibrium; 
furthermore the Ward identities associated to SUSY imply an equilibrium
 dynamics and in particular the fluctuation-dissipation theorem. 
The dynamical phase transition of mean field spin glass models is 
associated to a  spontaneously SUSY-breaking down to the subgroup of 
translations with respect to $\overline{\theta}$ \cite{kurfranz}. The 
initial conditions play for SUSY the same role as space boundary conditions in
 ordinary symmetry breaking. Therefore, different choices of initial 
conditions can lead to different asymptotic behaviours \cite{leticia,Franz,Alain}.\\
 In superspace the dynamical theory looks as a static theory for a
superfield with an internal coordinate ($t,\theta,
\overline{\theta}$) and a Hamiltonian $H$ with an extra quadratic term.
This similarity with a static theory allows to generalise the 
expansion in powers of $\beta$ described in section I to the dynamical case.
%%%%%%%%%%%%%%%%%%%%%%%%%%%%%%%%%%%%%%%%%%%%%%%%%%%%%%%%%%%%%%%%%%%%%
%  tap dinamiche
%%%%%%%%%%%%%%%%%%%%%%%%%%%%%%%%%%%%%%%%%%%%%%%%%%%%%%%%%%%%%%%%%%%%%
\subsection{Dynamical TAP equations}
In the following we focus on the Langevin dynamics of the p-spin spherical
 model for times not diverging with $N$. 
Within the static TAP approach one obtains a set of closed equations 
which, given
 the magnetic fields $h_{i}$ and fixed 
the spherical condition ($l=1$), have to 
be solved with respect to the local magnetisations $m_{i}$ 
and the spherical multiplier
 $\lambda$. In the dynamical case one has to compute a set of 
closed equations
 which, given the magnetic fields $h_{i}(t)$ and fixed the spherical condition 
$C(t,t)=1$, have to be solved with respect to the local 
magnetisations $m_{i}(t)$, the spherical 
multiplier $\lambda (t)$, the correlation and the response functions. 
This is the natural generalisation of the static TAP approach
 described in section I. As its static counterpart it allows 
to reconstruct the dynamical
 TAP free energy from a finite number of terms of the perturbative
 expansion.\\
To obtain the TAP dynamical free energy we apply the method of section I 
to the logarithm of the functional (\ref{superaction}): 
so we Legendre transform with respect
 to the 
(super)magnetisations $M_{i}(1)=\left<S_{i}(1)\right>$ and the two point 
 function $C(1,2)=\frac{1}{N}\sum_{i=1}^N \left<S_{i}(1)S_{i}(2)\right>$. The physical
 quantities $m_{i}(t)$, $C(t,t')$ and $R(t,t')$ are all encoded in these 
superspace functions.\\
The dynamical TAP free energy is:
\begin{equation}\label{gammadin}
-\beta \Gamma_{D} = \ln \int \prod_{i=1}^{N} {\cal D}S_{i} \exp (-{\cal L}'-
{\cal L} '')
\end{equation}
\[
{\cal L}'=-\beta \int d1  \sum_{1\leq i_{1}<\cdots <i_{p}\leq N} J_{i_{1},\cdots,i_{p}}S_{i_{1}}(1)\cdots S_{i_{p}}(1)+\frac{1}{2}\sum_{i=1}^{N}\int d1 d2 S_{i}(1)\Delta
(1,2)S_{i}(2) 
\]
\[
{\cal L}''=\int d1 \sum_{i=1}^{N}
 H_{i}(1)(S_{i}(1)-M_{i}(1))+ \frac{1}{2}\sum_{i=1}^{N}\int d1 d2 \Lambda (1,2)
(S_{i}(1)S_{i}(2)-C(1,2))\quad ,
\] 
where $\Delta(1,2)=D_{(1)}\delta(1-2)$ and  the delta function is defined 
 in the following way:
\[
 \delta(1-2)=\delta(t_{1}-t_{2})\delta(\overline{\theta}_{1}-\overline{\theta}_{2})\delta(\theta_{1}-\theta_{2})=\delta(t_{1}-t_{2})(\overline{\theta}_{1}-\overline{\theta}_{2})(\theta_{1}-\theta_{2})\quad .
\]
The Lagrange parameters $H_{i}(1)$ and $\Lambda(1,2)$ fix respectively the
 (super)magnetisation at each site $i$: $\left<S_{i}(1)\right>=M_{i}(1)$ and the 
two point function $C(1,2)=\frac{1}{N}\sum_{i=1}^{N}\left<S_{i}(1)S_{i}(2)\right>$.
\\
As we noted in the previous paragraph it is very important to specify
 the initial conditions. We take 
 as initial condition a fixed configuration $\{ s_{i}^{0} \}$; it does 
not matter if $\{ s_{i}^{0} \}$ is correlated or
 not with the couplings, because we do not 
average over disorder.
In appendix A we show how to take into account the
 initial condition
 in the derivation of the dynamical TAP equation.   
For the sake of clarity in the 
following this problem will not be addressed. \\ 
 As in section I, we construct $\Gamma _{D}$ through its power expansion in $\beta$ around $\beta=0$:
\begin{equation}\label{expgammadin}
-\beta \Gamma _{D}(\beta , M_{i},C)=\sum_{n=0}^{+\infty}
-\frac{\partial ^{n} 
(\beta \Gamma_{D})}{\partial \beta ^{n}}{\Bigg \vert}_{\beta = 0}
\frac{\beta ^{n}}{n!}\quad .
\end{equation}
The dynamical TAP equations are the Lagrange relations obtained from 
$\Gamma _{D}$:
\begin{equation}\label{lrel}
-\frac{\delta \beta \Gamma _{D}}
{\delta M_{i}(1)}=H_{i}(1)\qquad,\qquad 
-\frac{2}{N}\frac{\delta \beta \Gamma _{D}}
{\delta C(1,2)}=\Lambda(1,2)\quad .
\end{equation}
 
Since for $\beta=0$ the dynamical theory is Gaussian, the zeroth order
 term of (\ref{expgammadin}) reduces to:
%so we can 
% calculate explicitly the zeroth-order term of equation (\ref{expgammadin}) 
%as a function of
% $H_{i}$ and $\Lambda $;  then by deriving it
%with respect to $H_{i}$ and $\Lambda$ we find the relations among $H_{i}$,
% $\Lambda$ and $M_{i}$, $C$; finally inverting these relations and 
%substituting in  $-(\beta \Gamma_{D}){\big \vert}_{\beta=0}$, we find that:
\begin{equation}\label{g0adin}
-(\beta \Gamma _{D}){\Bigg \vert}_{\beta=0}=\frac{N}{2} Tr [ \ln (
 C-Q)] -\frac{N}{2}\int d1 d2 \Delta(1,2)C(1,2)\quad , 
\end{equation}
where $Q(1,2)=\frac{1}{N}\sum_{i=1}^{N}M_{i}(1)M_{i}(2)$ is the (super)overlap function.
The linear term in the right hand side of (\ref{expgammadin}) 
is a straightforward
 generalisation of its static counterpart (\ref{g1a}):
\begin{eqnarray}\label{g1adin}
-\beta \frac{\partial (\beta\Gamma _{D})}{\partial \beta} {\Bigg \vert}
_{\beta =0}&=&\beta \int d1 \sum_{1\leq i_{1}<\cdots <i_{p}\leq N} J_{i_{1}
,\cdots,i_{p}} \left<S_{i_{1}}(1)\cdots S_{i_{p}}(1)\right>_{\beta=0}\\
&&-\beta \int d1 \sum_{i}^{N} 
\frac{\partial H_{i}(1)}{\partial \beta}{\Bigg \vert}
_{\beta =0} \left<S_{i}(1)-M_{i}(1)\right>_{\beta=0}\nonumber\\
&&-\beta \int d1 d2 
\frac{\partial \Lambda(1,2)}{\partial \beta}{\Bigg \vert}
_{\beta =0} \sum_{i=1}^N \left<S_{i}(1)S_{i}(2)-C(1,2)\right>_{\beta=0}\quad .\nonumber
\end{eqnarray}
 As in section I the Lagrange conditions imply that the last two terms
 are zero. Since spins are decoupled at $\beta=0$, (\ref{g1adin}) simplifies 
to:
 \begin{equation}\label{g1bdin}
-\beta \frac{\partial (\beta\Gamma _{D})}{\partial \beta} {\Bigg \vert}
_{\beta =0}=\beta \int d1 \sum_{1\leq i_{1}<\cdots <i_{p}\leq N} J_{i_{1}
,\cdots,i_{p}} M_{i_{1}}(1)\cdots M_{i_{p}}(1)\quad .
\end{equation}
The analogy with the static case is evident also for the reaction term:
\begin{eqnarray}\label{g2adin}
 -\frac{\beta ^{2}}{2} \frac{\partial ^{2}(\beta\Gamma _{D})}{\partial 
\beta ^{2}} {\Bigg \vert}
_{\beta =0}  &=&  \frac{\beta ^{2}}{2}\left< \left(\sum_{1\leq i_{1}<\cdots <i_{p}\leq N} 
Y_{i_{1},\cdots,i_{p}}\right)^{2}\right>^{c}_{\beta =0}\quad ,\\
Y_{i_{1},\cdots,i_{p}} &=& \int d1{\Big [} J_{i_{1},\cdots,i_{p}}S_{i_{1}}(1)\cdots
S_{i_{p}}(1)- {\Big (}S_{i_{1}}(1)-M_{i_{1}}(1){\Big )}M_{i_{2}}(1)\cdots M_{i_{p}}(1)-\nonumber \\
&&- \cdots -M_{i_{1}}(1)\cdots 
M_{i_{p-1}}(1){\Big (}S_{i_{p}}(1)-M_{i_{p}}(1){\Big )}{\Big ]}\quad 
.\nonumber \\
\end{eqnarray}
To obtain the equation (\ref{g2adin}) 
we have used the following Maxwell relations:
\begin{eqnarray}\label{maxwelldyn}
\frac{\partial H_{i}(1)}{\partial \beta}{\Bigg \vert}_{\beta =0} & = & 
-\frac{\delta}
{\delta M_{i}(1)}\frac{\partial (\beta\Gamma _{D})}{\partial \beta}
{\Bigg \vert}_{\beta =0}\\
\frac{\partial \Lambda (1,2)}{\partial \beta}{\Bigg \vert}_{\beta =0} & = & 
-\frac{2}{N}\frac{\delta}
{\delta C(1,2)}\frac{\partial (\beta\Gamma _{D})}{\partial \beta}
{\Bigg \vert}_{\beta =0}\quad .
\end{eqnarray}
Using the statistical properties of the couplings it is easy to 
check that only the terms which correspond to squares of 
$J_{i_{1},i_{2}\cdots,i_{p}}$ give a contribution of the order of N  
and that the reaction term depends on 
the two point function $C(1,2)$ and the (super)overlap 
$Q(1,2)$ only.  Therefore we find
that the dynamical reaction term reduces to: 
\begin{equation}\label{g2cdin}
-\frac{\beta ^{2}}{2} \frac{\partial ^{2}(\beta\Gamma _{D})}{\partial 
\beta ^{2}} {\Bigg \vert}
_{\beta =0}  = \frac{\beta ^{2} N}{4}\int d1 d2 {\Big [}
 C(1,2)^p-Q(1,2)^p-p\left(C(1,2)-Q(1,2)\right)Q(1,2)^{p-1} {\Big ]} 
\quad .\nonumber
\end{equation}
As in
section I higher derivatives in equation 
(\ref{expgammadin}) can be neglected 
because they do not give a contribution of the order of N; so collecting expressions
 (\ref{g0adin}),(\ref{g1bdin}) and (\ref{g2cdin}) we find the dynamical TAP
 free energy:
\begin{eqnarray}\label{gfindin}
-(\beta \Gamma _{D})&=&\frac{N}{2} Tr [\ln(
 C-Q) ] -\frac{N}{2}\int d1 d2
 \Delta(1,2)C(1,2) +\\
&&+\beta \int d1 \sum_{1\leq i_{1}<\cdots <i_{p}\leq N} J_{i_{1}
,\cdots,i_{p}} M_{i_{1}}(1)\cdots M_{i_{p}}(1)+\nonumber \\
&&+\frac{\beta ^{2} N}{4}\int d1 d2 {\Big [}
 (p-1)Q(1,2)^p-pC(1,2)Q(1,2)^{p-1}+C(1,2)^p {\Big ]} \quad .\nonumber
 \end{eqnarray}
%We note that the method used to derived static TAP free energy can 
%be applied in a straightforward way to dynamics, because dynamics,
% in its superspace version, is nothing else that a "vector" generalization
% of the static theory. 
The dynamical TAP equations are obtained by the two Lagrange 
relations (\ref{lrel}):
\begin{equation}\label{dintapeq1}
\delta(1-3)=D_{1}{\Big (} C(1,3)-Q(1,3){\Big )}
+\int d2 {\Big [}\Lambda (1,2) -\mu \left(C(1,2)^{p-1}-Q(1,2)^{p-1}\right)
{\Big ]} {\Big (}C(2,3)-Q(2,3){\Big )}\quad ,\nonumber
\end{equation}
\begin{eqnarray}\label{dintapeq2}
D_{1}M_{i}(1)&+&\int d2 \Lambda (1,2) M_{i}(2)=-H_{i}(1)+
\sum _{1\leq i_{2}< \cdots  <i_{p}\leq N}  ' J_{i,i_{2}
,\cdots,i_{p}} M_{i_{2}}(1)\cdots M_{i_{p}}(1)-\\
&&-\mu \int d2 {\Big [} (p-1)\left(C(1,2)-Q(1,2)\right) Q(1,2)^{p-2}
-\left(C(1,2)^{p-1}-Q(1,2)^{p-1}\right) {\Big ]} M_{i}(2)\quad ,\nonumber
 \end{eqnarray}
where $\mu =\frac{p \beta ^{2}}{2}$ and the prime means 
that the sum does not run over $i$.
Note that the last term of (\ref{dintapeq2}) vanishes 
for $p=2$. This is natural because for $p=2$ the dynamical theory is 
Gaussian.
%%%%%%%%%%%%%%%%%%%%%%%%%%%%%%%%%%%%%%%%%%%%%%%%%%%%%%%%%%%%%%%%%%%%%%%%%%%
\subsection{Solution respecting causality}
{\it A priori} the dynamical TAP equations (\ref{dintapeq1}) and 
(\ref{dintapeq2}) may admit many different solutions, which can 
break partially or completely the symmetries of the action 
(\ref{action}). Up to now only the solution 
respecting causality\footnote{There is only one solution respecting
causality.}\cite{somzip} has been studied. The others 
 solutions are instantons, related to barrier crossing \cite{ioffe}.
In the following we focus on 
the causal solution, 
which breaks only partially the invariance
 of the action (\ref{superaction}) \cite{kurfranz}. The unbroken symmetries  
 allow to find the general form of this solution \cite{kurfranz}:
\begin{equation}\label{M}
M_{i}(1)=m_{i}(t)
\end{equation}
 \begin{equation}\label{C}
C(1,2)=C(t_{1},t_{2})+(\overline{\theta}_{1}-\overline{\theta}_{2})(\theta_{1}R(t_{2},t_1)-
\theta_2 R(t_1,t_2))\quad .
\end{equation}
The spherical constraint $C(1,1)=1$
 is fixed through $\Lambda (1,2)$, which has the usual form\cite{kurfranz}:
\begin{equation}
\Lambda(1,2)=\delta(1-2)\lambda(t_{1})\quad ,
\end{equation}
where $\lambda(t)$ is a real function of time.\\
Plugging these expressions in the dynamical TAP equations (\ref{dintapeq1})
 and (\ref{dintapeq2}),
 we find that the magnetisations and the correlation and the
 response functions satisfy for $t>0$ and $t'>0$ the following equations:
\begin{eqnarray}\label{tc}
\frac{\partial}{\partial t}{\Big (} C(t,t')-Q(t,t'){\Big )} &=&2R(t',t)-\lambda(t)
{\Big (}C(t,t')-Q(t,t'){\Big )}+\\
&&+\mu \int_{0}^{t'}dt''{\Big (}C(t,t'')^{p-1}-Q(t,t'')^{p-1}{\Big )}
R(t',t'')\nonumber \\
&&+ \mu (p-1)\int_{0}^{t}dt''{\Big (}C(t'',t')-Q(t'',t'){\Big )}R(t,t'')C(t,t'')^{p-2}\quad ,\nonumber
\end{eqnarray}
\begin{eqnarray}\label{tr}
\frac{\partial}{\partial t}R(t,t')&=&-\lambda(t)R(t,t')+\delta(t-t')+\\
&&+\mu (p-1)\int _{t'}^{t} dt''R(t,t'')R(t'',t')C(t,t'')^{p-2}\quad ,\nonumber
\end{eqnarray}
\begin{eqnarray}\label{tm}
\left(\frac{\partial}{\partial t}+\lambda (t)\right)m_{i}(t)&=&
\beta h_{i}(t)+\beta 
\sum_{1\leq i_{2}< \dots <i_{p}\leq  N }'J_{i,i_{2},\dots ,i_{p}}
m_{i_{2}}(t)\cdots m_{i_{p}}(t)+\\
&&+\mu (p-1) \int_{0}^{t} dt'' {\Big (}C(t,t'')^{p-2}-Q(t,t'')^{p-2}){\Big )}
R(t,t'')m_{i}(t'')\quad ,\nonumber
\end{eqnarray}
where $Q(t,t')=\frac{1}{N}\sum_{i=1}^{N}m_{i}(t)m_{i}(t')$ is the overlap 
function and $h_{i}(t)$ is the magnetic field acting on the {\em i}th spin.
 The correlation function satisfies the boundary 
condition $C(t,0)=Q(t,0)$ and 
 magnetisations fulfil the initial
 conditions $m_{i}(0)=s_{i}^{0}$ (see appendix A).\\
 Moreover the spherical
 condition $C(t,t)=1$ fixes $\lambda $ as a function of time through the
 equation:
\begin{eqnarray}\label{lambda}
\lambda (t){\Big (}1-q(t){\Big )}&=&1+\frac{1}{2}\frac{dq}{dt}+\mu 
\int_{0}^{t}dt''{\Big (}C(t,t'')^{p-1}-Q(t,t'')^{p-1}{\Big )}R(t,t'')+\\
&&+\mu (p-1) \int_{0}^{t}dt''{\Big (}C(t'',t)-Q(t'',t){\Big )}R(t,t'')C(t,t'')^{p-2}\quad ,\nonumber
\end{eqnarray}
where $q(t)=Q(t,t)$.\\
In the following we show that starting from 
(\ref{tc}), (\ref{tr}), (\ref{tm}) and (\ref{lambda}) one can obtain 
 the dynamical equations of \cite{leticia} as a particular case.
Actually, if one takes as initial condition for the dynamical measure 
an uniform average  
over all possible configurations as in\cite{leticia},
 then the magnetisations are equal to zero at $t=0$ and there 
is no boundary condition on the correlation function (see appendix A). 
In this case we find that the equation (\ref{tm}) is trivially satisfied and
 the equations 
(\ref{tc}), (\ref{tr}) and (\ref{lambda}) reduce 
to the ones considered in \cite{leticia}. \\
Furthermore it is easy to verify 
that at zero temperature (\ref{tm})
 coincides with a simple gradient descent, as should be because the 
 thermal noise is absent; to handle the zero temperature limit  
we write (\ref{tm}) in terms of the real time 
$t_{r}=\beta t$, finding
\begin{eqnarray}\label{tmrealtime}
\left(\frac{\partial}{\partial t_{r}}+\overline{\lambda}  (t_{r})\right)m_{i}(t_{r})&=& h_{i}(t_{r})+
\sum_{1\leq i_{2}< \dots <i_{p}\leq  N }'J_{i,i_{2},\dots ,i_{p}}
m_{i_{2}}(t_{r})\cdots m_{i_{p}}(t_{r})+\\
&&+\frac{p(p-1)}{2} \int_{0}^{t_{r}} dt_{r}'' \left(C(t_{r},t_{r}'')^{p-2}
-Q(t_{r},t_{r}'')^{p-2})\right)
R(t_{r},t_{r}'')m_{i}(t_{r}'')\quad , 
\end{eqnarray}
where $\overline{\lambda}= \frac{\lambda }{\beta }$. 
For $T=\frac{1}{\beta }=0$ the 
last term in equation (\ref{tmrealtime}) is zero because 
 without thermal noise $Q(t_{r},t_{r}')=C(t_{r},t_{r}')$.
Therefore
we recover the zero temperature limit of the Langevin equations: a pure
 gradient descent in the energy landscape.

In summary for the solution respecting causality we have derived the 
dynamical TAP equations (\ref{tc}), (\ref{tr}), (\ref{tm}) and (\ref{lambda}). These 
 equations on the correlation and the 
response functions and on the local magnetisations
 have to be solved for a given realization of the couplings and for
 a given initial condition. \\
Note that the equations on local magnetisations do not
 have the form of a gradient descent in the free energy landscape because
 the Onsager reaction term is non-Markovian.
 This is natural because it represents the contribution 
to the effective field of the $i$th spin 
 due to the influence at previous times 
of the $i$th spin on the others.
%%%%%%%%%%%%%%%%%%%%%%%%%%%%%%%%%%%%%%%%%%%%%%%%%%%%%%%%%%%%%%%%%%%%%
%  soluzioni asintotiche
%%%%%%%%%%%%%%%%%%%%%%%%%%%%%%%%%%%%%%%%%%%%%%%%%%%%%%%%%%%%%%%%%%%%%
\section{Asymptotic analysis}
In the following we perform an asymptotic analysis of the equations
 (\ref{tc}), (\ref{tr}), (\ref{tm}) and (\ref{lambda}). For the sake of
 simplicity we will take $h_{i}(t)=0$ in (\ref{tm}). 

Two asymptotic behaviour have been found for the p-spin spherical model
depending on the choice of the initial conditions
\cite{leticia,Franz,Alain}:
\begin{itemize}
\item True ergodicity breaking: the system equilibrates 
 in a separate ergodic component. Asymptotically time homogeneity 
 and fluctuation-dissipation theorem (FDT) hold\cite{Franz,Alain}. 
\item Weak ergodicity breaking\footnote{The concept of weak ergodicity breaking was introduced in \cite{Bouchaud}.}: the system does not equilibrate.
 Asymptotically two time sectors can be identified. 
 In the first one (FDT regime),
 which corresponds to finite time differences $|t-t'| \sim O(1)$, 
 ($t>>1$, $t'>>1$),
 the system has a pseudo-equilibrium dynamics since
 FDT and time translation invariance hold asymptotically. 
In the second one (ageing regime), which corresponds
 to ``infinite'' time differences $|t-t'|\sim t'$, 
FDT and time translation invariance do not apply and the system
 ages\cite{leticia}.
\end{itemize}
These two dynamical behaviours correspond to different Ans{\"a}tze for 
the asymptotic form of the two time quantities. Following \cite{Franz,Alain} 
 we take for the equilibrium dynamics in a separate ergodic
 component the Ansatz ($t>t'>>1$): 
\begin{eqnarray}\label{eqansatz}
C(t,t')=C_{FDT}(t-t')&\qquad ,\qquad &R(t,t')=R_{FDT}(t-t')\\
R_{FDT}(\tau )=-\theta(\tau )\frac{dC_{FDT}(\tau )}{d\tau }&\qquad ,\qquad &
Q(t,t')=q\\
\lim_{\tau \rightarrow \infty }C_{FDT}(\tau )=q.&\qquad \qquad  &
\end{eqnarray}
In the case of slow dynamics we take for finite time separations the
 Ansatz corresponding to equilibrium dynamics, but with $Q(t',t)=q'$. The 
 difference between $q$ and $q'$ already indicates
 that the dynamics is not characterised by a true breaking of ergodicity and
 that the system does not equilibrate in a separate ergodic component. Whereas
 for the ageing sector we take the Ansatz\footnote{The asymptotic
 equations are obtained neglecting the time derivatives. This has as
 a consequence that from an asymptotic solution we obtain infinitely 
 many others
 by re-parameterisation \cite{leticia}. For the sake of clarity 
in the following we focus on the particular
 parameterisation shown in equations (\ref{slansatz1}), (\ref{slansatz2})
 and (\ref{slansatz3}) . } \cite{leticia}:
\begin{eqnarray}\label{slansatz1}
C(t,t')=q C_{ag}(\lambda )&\qquad ,\qquad &t R(t,t')=R_{ag}(\lambda )\\
R_{ag}(\lambda )= x q \frac{d C}{d \lambda }&\qquad ,\qquad &
Q(t,t')=q' Q_{ag}(\lambda )\label{slansatz2}\\
C_{ag}(1)=Q_{ag}(1)=1 &\qquad ,\qquad & \lambda =\frac{t'}{t},\label{slansatz3}
\end{eqnarray}
where $x$ parameterises the violation of FDT. Note that in the usual case
the overlap function is not present.\\
In the next sections we will follow this strategy: assuming some properties
 on the dynamical evolution we will analyse the 
 asymptotic solutions which are consistent with this assumption;
 then following the physical picture associated to each asymptotic solution
 we will propose which are the initial conditions related to 
 this asymptotic solution. \\
The matching between asymptotic behaviour 
 and initial conditions is a general problem in mean field spin glass
 dynamics; up to now there are no analytical methods available and
 one has to resort to numerical integration of the dynamical equations. 
 Nevertheless in our case the integro-differential character of the dynamical
 equations and their number (2 equations on two time quantities and N
 equations on one time quantities) makes difficult to reach very
 long times. 
%%%%%%%%%%%%%%%%%%%%%%%%%%%%%%%%%%%%%%%%%%%%%%%%%%%%%%%%%%%%%%%%%%%%%%
%  dinamica di equilibrio
%%%%%%%%%%%%%%%%%%%%%%%%%%%%%%%%%%%%%%%%%%%%%%%%%%%%%%%%%%%%%%%%%%%%%%
\subsection{Equilibrium dynamics}\label{eqdyn}
 In this section
 we assume that the system has an equilibrium dynamics and we analyse all the 
 asymptotic solutions which are consistent with this assumption.\\
 We denote respectively by 
$\lambda ^{\infty }$ and $m_{i}^{\infty }$ the
 asymptotic values of the spherical multiplier and 
of the local magnetisations. Plugging the equilibrium dynamics Ansatz
 into (\ref{tm}) and (\ref{lambda})
 we find that the equations on $m_{i}^{\infty }$ and 
$\lambda ^{\infty }$ are 
the static TAP equations (\ref{tapstatichem}) and (\ref{tapstatichel}).
 In the asymptotic limit the equations (\ref{tc}) and (\ref{tr}) 
on the correlation
 and the response functions reduce to: 
\begin{equation}\label{eqtapc}
 \left(\frac{d}{d \tau }+\lambda ^{\infty }-\mu \right)C(\tau)+
\mu +1-\lambda ^{\infty }=-\mu \int_{0}^{\tau }d\tau ' C(\tau  -\tau ')^{p-1}
 \frac{dC(\tau ')}{d\tau '}\quad .
\end{equation}
The above equation describes the equilibrium dynamics inside the ergodic 
component associated to a TAP solution $\{m_{i}^{\infty} \}$. Note
 that this asymptotic dynamical 
solution is consistent with the assumption of an 
equilibrium dynamics 
 only if $\{m_{i}^{\infty} \}$ is 
 a local minimum of the free energy. \\
As we have remarked before, the asymptotic 
analysis does not give any information about the relationship among initial
 conditions and asymptotic solutions. However, since 
 this asymptotic solution represents the
equilibration in a stable TAP state $\{m_{i}^{\infty} \}$, it
 is quite natural to associate to this solution
 an initial condition belonging to this state.
This interpretation is suggested by the results of \cite{Franz,Alain}.
 Indeed in \cite{Franz,Alain} 
the low temperature dynamics has been studied 
 starting from an initial condition thermalized at a temperature $T'$ between 
 the statical and the dynamical transition temperatures. This 
procedure corresponds to take an
initial condition belonging to the TAP states, which are the 
equilibrium states at the temperature $T'$. 
In \cite{Franz,Alain} it has been shown
 that the system relaxes in the 
 TAP states associated to the initial condition. It is easy to show
 that the equation 
 satisfied by $C(\tau )$ in \cite{Franz,Alain} can be written in the form
(\ref{eqtapc}).\\
Besides, to elucidate how dynamical quantities
 approach their asymptotic values, we write
 equation (\ref{tm}) in a more appealing form:
\begin{eqnarray}\label{gradiente}
\frac{dm_{i}}{dt}&=& -\beta \frac{\partial \Gamma_{S} }{\partial m_{i}}+f(t)m_{i}
 +\beta h_{i}'(t)\quad ,\\
f(t)&=&\frac{1}{1-q}+\mu (1-q^{p-1})-\lambda (t)\quad ,\nonumber \\
\beta h_{i}'(t)&=&-\mu   \left(1-q^{p-1}-(p-1)q^{p-2}(1-q)
 \right)m_{i}+\nonumber \\
&&+\mu 
(p-1) \int _{0}^{t}dt'' \left(C(t,t'')^{p-2}-Q(t,t'')^{p-2})\right)
R(t,t'')m_{i}(t'') \quad ,\nonumber 
\end{eqnarray}
where $\Gamma_{S}$ is the static TAP free energy and 
$f(t)$ and $h_{i}'(t)$ are
 functions going to zero at large times. \\
The equations (\ref{tm}) 
  show that the dynamical evolution of magnetisations
 looks like a gradient descent in the free energy landscape but with 
 an extra spherical multiplier $f(t)$ and magnetic fields $h_{i}'(t)$  
 correlated with the initial condition. As it should be for an
equilibrium dynamics, all 
  quantities are characterised by  an exponential relaxation (to zero
 for $f(t)$ and $h_{i}'(t)$), therefore 
 it is not possible to identify any slow or fast variables and the reaction 
 term remains non-Markovian at all time (except $t=\infty $). 
  Anyway the interpretation of 
 this asymptotic solution as an equilibration in a TAP state 
 shows up explicitly from (\ref{gradiente}). 

In summary, for an asymptotic solution corresponding to an equilibration in
 a stable TAP state the dynamical probability
 measure relaxes exponentially fast toward the equilibrium measure associated
 to this state and at large times the evolution of the local magnetisations
 looks like a gradient descent in the free energy landscape, but with 
 the extra spherical multiplier $f(t)$ and the magnetic fields $h_{i}'(t)$
 going to zero.
%%%%%%%%%%%%%%%%%%%%%%%%%%%%%%%%%%%%%%%%%%%%%%%%%%%%%%%%%%%%%%%%%%%%%%%%%%%
%  Dinamica di non equilibrio
%%%%%%%%%%%%%%%%%%%%%%%%%%%%%%%%%%%%%%%%%%%%%%%%%%%%%%%%%%%%%%%%%%%%%%%%%%%
\subsection{Non equilibrium dynamics}
 %%%%%%%%%%%%%%%%%%%%%%%%%%%%%%%%%%%%%%%%%%%%%%%%%%%%%%%%%%%%%%%%%%%%%%%%%%
% C.I. random
%%%%%%%%%%%%%%%%%%%%%%%%%%%%%%%%%%%%%%%%%%%%%%%%%%%%%%%%%%%%%%%%%%%%%%%%%%
\subsubsection{Weak ergodicity breaking}
 In the following
 we assume that the system has a 
 slow dynamics with $q\neq q'$. As we already pointed 
 out the difference between $q$ and $q'$ marks that the system does not 
 equilibrate in a single ergodic component. \\
 The asymptotic analysis in the time sector corresponding to finite 
 time differences leads to the same equation (\ref{eqtapc}) 
 for the correlation and the response functions. Whereas for infinite time
 differences we find 
  that the asymptotic equations admit the solution: $q'=0$, $q$ 
which coicides with the overlap 
of the threshold states \cite{ptap,crisantitap},
 $x=\frac{(p-2)(1-q)}{q}$ and $C_{ag}(\lambda)$ and $R_{ag}(\lambda)$, which
 satisfy the same equations found in \cite{leticia}.\\
The equation (\ref{lambda}) on the spherical multiplier reduces to:
 $\lambda ^{\infty }=(1-q)^{-1}+\mu (1-q^{p-1})$ 
 and the asymptotic value of the local magnetisations 
 $m_{i}^{\infty }$ is zero. This is exactly the same asymptotic solution
 found in \cite{leticia} for random initial conditions. 
 Therefore it is natural to associate to this solution a random initial
 condition, which is not correlated with any particular stable TAP
 state.\\
  To study how the local magnetisations vanish at large times,
 the ``mean field'' energy term in  
 (\ref{tm}) can be neglected because is of the order of 
$m_{i}^{p-1}$. Moreover we can substitute 
 to all terms which 
 multiply $m_{i}$ their asymptotic values, finding that (\ref{tm}) reduces to:
\begin{equation}\label{tapmas}
\left(\frac{\partial}{\partial t}+\lambda ^{\infty}\right)m_{i}(t)=
\mu (p-1) \int_{0}^{t} dt'' C(t,t'')^{p-2}
R(t,t'')m_{i}(t'')\quad .
\end{equation}
This is exactly the same equation satisfied by $C(t,t')$ and $R(t,t')$ 
 for a fixed and finite $t'$ and a very large value of $t$ \cite{leticia}.
 As a consequence the local magnetisations and (for a fixed $t'$)
 the correlation and the
response functions go to zero 
 at large times in the same way and
 because of the same mechanism: fixed any configuration
 reached during the dynamics, two different typical noise histories
 bring the system in two completely uncorrelated configurations \cite{Alain2}.

Besides, we remark that the local magnetisations vanish at large time, whereas
 the correlation function tends in the FDT regime to 
 the overlap of the threshold states. This seems to indicate that
 the dynamical measure tends toward (for the one time quantity)
 a static measure which is broken in separate components, i.e. the
 threshold states. In other words and more formally the fact that
 the overlap function $Q(t,t')$ tends to zero and the correlation function
 tends (in the FDT regime) to the overlap of the threshold states implies
 that the dynamical probability measure is not a ``pure'' probability
 measure, because the clustering in time is not valid in the FDT regime. 
\subsubsection{Between true and weak ergodicity breaking}
 In the following we assume that the system has a 
slow dynamics such that 
$\lim_{t\rightarrow \infty }
 C_{FDT}(t)=\lim_{t\rightarrow \infty }1/N\sum_{i=1}^{N}m_{i}^{2}(t)$,
 i.e. $q=q'$.\\
 The asymptotic analysis in the time sector corresponding to finite 
 time differences leads to the same equation (\ref{eqtapc}) 
 for the correlation and the 
response functions. Whereas for the ``infinite'' time
 difference sector ($|t-t'|\sim t'$) we find 
 that the asymptotic equations admit the solution:
 \begin{equation}\label{sol}
 Q_{ag}(\lambda)=C_{ag}(\lambda)  \qquad x=\frac{(p-2)(1-q)}{q}\quad ,
 \end{equation}
 where $R_{ag}(\lambda )$ and $C_{ag}(\lambda)$ 
satisfy the same equations 
 found in \cite{leticia}, $q=q'$ satisfies the equation for the 
 overlap of the threshold states \cite{ptap,crisantitap}:
\begin{equation}\label{theq}
\frac{1}{p-1}=\mu q^{p-2} (1-q)^{2}\quad ,
\end{equation}
 and the equation (\ref{lambda}) on the spherical 
 multiplier reduces to (\ref{tapstatichel}). The previous results,
 in particular that the correlation and response functions
 fulfil the equation of the equilibrium relaxation dynamics inside 
a threshold state and the equality between 
$q'$ and $q_{threshold}$, indicate that at very large times  
the system has almost thermalized within 
 a threshold 
 state.\\
Note that the asymptotic form
 of (\ref{tc}) is automatically verified for
 $Q_{ag}(\lambda)=C_{ag}(\lambda)$; whereas (\ref{tr}) coincides 
 in the asymptotic limit with the the first one of the two coupled equations on
 $C_{ag}(\lambda)$ and $R_{ag}(\lambda )$ found in \cite{leticia}. Another
 equation is obtained applying to (\ref{tm})
 the Martin-Siggia-Rose approach. This 
allows to average over the couplings
 and to obtain another equation on $Q_{ag}(\lambda)$ and $R_{ag}(\lambda )$.  
As expected, if one takes $Q_{ag}(\lambda)=C_{ag}(\lambda)$
 this equation coincides with the second one verified by $C_{ag}(\lambda)$ in 
\cite{leticia}. \\
Therefore we have found that this asymptotic solution 
is very similar to the one associated to random initial conditions. The
only difference is that for the latter the local
 magnetisations vanish because of the 
spreading of the dynamical probability measure at finite times. As we 
have suggested in the 
previous section, it seems that the dynamical probability measure (starting
 from random initial conditions) tends toward
 a static probability measure which is broken in separate components, 
i.e. the threshold states. Whereas for the asymptotic
 solution studied in this section the effects due to the spreading of the
 dynamical probability measure are absent and the system seems to equilibrate
 (in the FDT time regime) within a threshold state. As a consequence
it seems natural that the initial conditions related to
 this asymptotic solution 
are the configurations typically reached in the long-time dynamics
 (starting from a random initial condition). \\
In fact a way to obtain this
 asymptotic solution starting from a random initial condition 
is to introduce fields $h_{i}(t)$ which enforce the
 condition $\lim_{t\rightarrow \infty }1/N\sum_{i=1^{N}}m_{i}(t)^{2}=q'
=q_{th}$.
There are many different way to fix the fields $h_{i}(t)$ 
to enforce this condition; however for each realization of $h_{i}(t)$ 
it is clear that 
 $lim_{t\rightarrow \infty }h_{i}(t)=0$, because 
 the condition $q'=q_{th}$ is automatically verified in the
 long-time limit.
 The spreading of the dynamical measure is due 
\cite{Alain2} to 
 the many possible paths that the system can follow in the energy landscape.
 A particular realization of the noise 
brings the system along a particular
 path. The role of the magnetic fields $h_{i}(t)$ is to avoid the spreading of the dynamical measure: a particular realization of $h_{i}(t)$  brings 
the system along one of the possible paths.\\
Recently Franz and Virasoro \cite{qe} have developed an interpretation
of the equality between the fluctuation dissipation ratio ($x$, see eq.
 (\ref{slansatz2})) and the rate of growth of the complexity
 close to the asymptotic state in terms of 
quasi-equilibrium concepts. In particular they have introduced the notion
 of quasi-equilibrium state. The local magnetisations of a quasi-equilibrium
 state are defined for a fixed thermal noise as \cite{qe}: $m_{i}^{q-eq}(t)=
1/\tau \int_{t-\tau }^{t}S_{i}(t')dt'$, where $\tau $ is such that
 $C(t,t-\tau )=q_{th}$. The analysis of the equations of motion 
\cite{preparation} for $m_{i}^{q-eq}(t)$ confirms and  
 strengthens the previous interpretation of $h_{i}(t)$. 

In the following we analyse the slow evolution of local magnetisations.
 Note that since the overlap and the correlation functions are equal
 in the ageing regime, 
 one can obtain the slow dynamical behaviour of the correlation
 and the response functions starting from the evolution of $m_{i}(t)$ .\\
 As in the previous section the equations on the local magnetisations
 can be rewritten as a gradient descent in the free energy landscape
 with the extra spherical multiplier $f(t)$ and the extra magnetic fields
 $h'_{i}(t)$ going to zero at large times. Therefore (\ref{gradiente}) 
 implies that 
\begin{equation}\label{limite}
\lim_{t\rightarrow \infty }\frac{\partial \Gamma_{S} }{\partial m_{i}}(m_{i}(t))=0\quad .
\end{equation}

Thus, even if the analysis of the FDT regime shows that 
at very large times the system has almost thermalized within 
 a threshold 
 state, the fact that the overlap function shows an ageing  
 behaviour implies that magnetisations do not tend toward a particular
 threshold state. In other words the local
 magnetisations continue to evolve forever (even if more and more slowly) 
 and their large time limit does not exist.
 This already happens for the spherical
 Sherrington-Kirkpatrick model, for which an  
 exact analytical solution is available \cite{Dean}. \\
As it is implicitly 
 contained in the slow dynamic assumption, the local 
 magnetisations are (asymptotically)
 constant on time-scales associated to the FDT regime 
 and their evolution is on time-scales associated to the ageing regime.
 Because of almost flat directions $f(t)$ and
 $h_{i}'(t)$ play a fundamental role and are responsible for ageing. 
 In fact at large times the dynamics takes place only along 
 almost flat directions and these vanishing function of time act as 
 a vanishing source of drift, so the larger is the time, the
 weaker is the drift and the slower is the evolution: the system
 ages. 
%%%%%%%%%%%%%%%%%%%%%%%%%%%%%%%%%%%%%%%%%%%%%%%%%%%%%%%%%%%%%%%%%%%%%%%%%%%%%
\subsection{Free energy landscape and long-time dynamics}
%%%%%%%%%%%%%%%%%%%%%%%%%%%%%%%%%%%%%%%%%%%%%%%%%%%%%%%%%%%%%%%%%%%%%%%%%%%%%
At finite times the dynamics cannot be represented as 
an evolution in the free energy landscape because the Onsager reaction term
 in (\ref{tm}) is non-Markovian. \\
Only in the asymptotic time regime it is possible
 to establish a connection between the free energy 
landscape and the dynamical evolution.\\
When one takes an initial condition which leads to an equilibrium dynamics,
 i.e. the equilibration in a stable TAP state $\{m_{i}^{\infty }\}$,
 the equations on the local magnetisations imply that the relaxation
 of $\{ m_{i}(t) \}$ toward $\{m_{i}^{\infty }\}$ coincides
  with a gradient descent in the free energy
 landscape  with an extra spherical multiplier and magnetic fields
 going to zero at large times.\\
However, in the most interesting and the most physical
 case of random initial conditions
the local magnetisations vanish at large times. Thus, the dynamical
 evolution is dominated by the threshold states, but the asymptotic
 evolution of the local magnetisations does not show any indication of this.\\
 This result can be understood thinking to a 
Langevin dynamics in a double
 well potential $V(x)$. 
At large times the dynamical probability measure is 
 equally distributed on the two wells, therefore the mean position
 $\left< x(t) \right>$ is zero. As a consequence 
 the mean position gives a very poor description
 of the probability measure; whereas the second moment 
$\left< x(t)x(t') \right>$ gives more insight into the probability
 distribution. The same thing happens for the p-spin spherical model
 for random initial conditions: the local magnetisations
 $\left< s_{i}(t) \right>$ vanish at large times and do not give a good
 representation of the asymptotic dynamical probability measure, whereas
 the two time quantities $C(t,t')$ and $R(t,t')$ do.\\
However, a description of the asymptotic dynamics as an evolution
 in the flat directions of the free energy landscape makes sense. 
To avoid the previous problem,
 i.e. the spreading of the dynamical measure, one can take
 as initial condition a configuration 
typically reached in the long-time dynamics\footnote{In the analogy with the 
double well potential this procedure is equivalent
 to take the initial condition in one of the two wells}. \\
In this case the correlation and the response functions 
have the same asymptotic behaviour that for a random initial condition. Moreover
 $C(t,t')$ and $Q(t,t')$ are equal in the ageing time regime.
 Thus, the ageing dynamics obtained
 starting from a random initial condition 
can be represented in terms of the equation
 (\ref{gradiente}), i.e. as a motion in 
the flat directions of the free energy landscape.\\
In summary we have found that the representation of the long-time
 dynamics as an evolution in the free energy landscape is
correct. This
evolution consists in a gradient descent in the free energy landscape
 with an extra spherical multiplier and extra magnetic fields
going to zero at large time. These vanishing sources of drift depend
on the history of the system and are crucial for slow dynamics. 
%%%%%%%%%%%%%%%%%%%%%%%%%%%%%%%%%%%%%%%%%%%%%%%%%%%%%%%%%%%%%%%%%%%%%%%%%%%%
% INM
%%%%%%%%%%%%%%%%%%%%%%%%%%%%%%%%%%%%%%%%%%%%%%%%%%%%%%%%%%%%%%%%%%%%%%%%%%%%
\section{Instantaneous Normal Modes analysis of the energy landscape}\label{INM}
In this section we analyse the local properties of the 
energy landscape seen by the system during the
dynamical evolution.
In particular we  carry out a computation of the spectrum of the
 energy Hessian for a typical
 dynamical configuration. The eigenvectors of the energy Hessian are 
called Instantaneous Normal Modes \cite{keyes} and 
 they have been introduced in liquid theory 
to represent the short time dynamics within 
 a harmonic description.

To be more specific, consider the energy Hessian of the p-spin spherical model:
\begin{eqnarray}\label{hessian}
H_{i,j}&=&-E_{i,j}+
\frac{\lambda }{\beta }\delta _{i,j}\\
E_{i,j}&=&\frac{p (p-1)}{p!}\sum_{i_{3}\neq \dots \neq i_{p}(\neq i,j)}J_{i,j,i_{3},\dots ,i_{p}}
s_{i_{3}}\cdots s_{i_{p}}
\end{eqnarray} 
and the density of states 
\begin{equation}\label{rho}
\rho (\mu;t)=\overline{\left<\sum_{\alpha =1}^{N} \delta (\mu -\mu_{\alpha })\right>}\quad , 
\end{equation}
where $\mu_{\alpha }$ is an eigenvalue of (\ref{hessian}), $\left< \cdot \right>$ means 
 the average over the dynamical configurations at time t 
and the over-line  indicates
 the average over the couplings. As we will show in the following, 
$\rho (\mu ;t)$ is a self-averaging quantity; therefore the typical 
 and the average value of $\rho (\mu ;t)$ are the same (in the large-N
 limit). \\
Now we present a sloppy derivation of $\rho (\mu;t)$, a more careful
 derivation is shown in appendix B.\\
Since the density of states of ${\mathbf E}$ and the spectrum of 
 the energy Hessian (\ref{hessian}) are related by 
 $\frac{\lambda (t)}{\beta }$ through a translation, in the following
 we focus on ${\mathbf E}$.\\
To compute 
 the correlation functions of the elements of ${\mathbf E}$ one can safely
 assume \cite{moore,laloux} that the configurations $s_{i}$
 are uncorrelated with the couplings at the leading order in $N$, finding:
\begin{eqnarray}\label{goe}
\overline{\left<E_{i,j}\right>}&=&0 \qquad \qquad i\leq j\\
\overline{\left<E_{i,j} E_{k,l}\right>}&=&\delta_{i,k} 
\delta_{j,l} \frac{b^{2}}{4}\qquad 
i\leq j,k \leq l.
\end{eqnarray}
Where $b^2$ is equal to
\begin{eqnarray}\label{b}
b^{2}&=&4\left[\frac{p (p-1)}{p!} \right]^{2}\sum_{i_{3}\neq  \dots \neq  i_{p}
(\neq i,j) }
\sum_{j_{3}\neq  \dots \neq  j_{p}(\neq i,j)}
\overline{J_{i,j,i_{3},\dots ,i_{p}}
J_{i,j,j_{3},\dots ,j_{p}}\left<s_{i_{3}}\cdots s_{i_{p}}s_{j_{3}}\cdots s_{j_{p}}\right>}\\
&=&4\left[\frac{p (p-1)}{p!} \right]^{2}N^{p-2}\left(\frac{p!}{2 N^{p-1}}\right)
(p-2)!C(t,t)^{p-2}=\frac{2 p(p-1)}{N} \qquad (i<j)\quad . 
\end{eqnarray}
The matrix ${\mathbf E}$ is nothing else that a Gaussian
 random matrix with variance $N^{2} \frac{b^{2}}{4}$. Therefore its typical
 (and average) density of
 states (see for example \cite{mehta}) is a semi-circle law centred in $0$
 with support $[-\sqrt{2p(p-1),2p(p-1)}]$. As a consequence the spectrum
 of the energy Hessian (\ref{hessian}) equals:
\begin{equation}\label{semicirc}
\rho (\mu;t)=\frac{1}{\pi p(p-1)}\sqrt{2p(p-1)
-\left( \mu -\frac{\lambda (t)}{\beta}\right)^{2}}\quad .
\end{equation} 
At large times $\lambda (t)$ converges to its limiting value 
$\lambda ^{\infty }$ and $\rho (\mu;t)$ converges to $\rho ^{\infty }(\mu)$,
 which is obtained replacing $\lambda ^{\infty }$ to $\lambda (t)$ 
in (\ref{semicirc}). 
In the following we consider the dynamics starting from a random initial
 condition. In this case $\lambda ^{\infty }$ verifies the following equation: 
\begin{equation}
\frac{\lambda^{\infty}}{\beta}=\frac{1}{\beta (1-q)}+\frac{p \beta }{2}
(1-q^{p-1})\quad ,
\end{equation}
where $q$ is equal to zero for $T>T_{d}$ and fulfils the equation (\ref{theq})
of the overlap of the
 threshold states for 
$T< T_{d}$, $T_{d}=\sqrt{\frac{p (p-2)^{p-2}}{2 (p-1)^{p-1}}}$.\\
In fig.1 the left edge ($\lambda _{min}$) 
of the spectrum $\rho ^{\infty }(\mu)$
 is plotted as a function of $\beta=1/T$ for $p=3$. 
At very high
 temperature all the eigenvalues are positive because the energy landscape
 seen by the system is dominated by the quadratic potential which fix
 the spherical constraint. In this regime decreasing the temperature
 the energy landscape seen by the system becomes more and more rugged and
 the minimum eigenvalue decreases and becomes negative. 
On the other hand at very low temperature there is a finite 
fraction of negative eigenvalues. In this case if the temperature
 decreases then
 the minimum eigenvalue of the energy Hessian grows and 
equals zero at zero temperature, as
 it is expected for a dynamics in a rugged energy landscape.\\
Therefore analysing the left edge of $\rho ^{\infty }(\mu)$
 we have found a crossover
 from a high temperature regime in which the system is substantially
 confined in a harmonic potential and a low temperature regime in which
 the ruggedness of the energy landscape seen by the system during the
 dynamical evolution becomes more and more relevant. We can interpret 
 the temperature  
 at which $\lambda _{min}$ reaches its
 minimum value as a crossover temperature 
$T_{0}=1/\beta _{0}=\sqrt{p/2}$ between these two temperature regimes.
 It is quite interesting to note that $T_{0}$ has a relationship
 with the damage spreading transition \cite{ritort}.\\
In fact it has been shown in \cite{ritort} 
that the p-spin spherical model exhibits
 a damage spreading transition at a temperature $T_{ds}$, which satisfies
 the inequality: $\sqrt{p/2-1}\leq T_{ds} <\sqrt{p/2}$. This
result shows that 
 the temperature $T_{0}$ arising in the study of local properties of the energy
 landscape is related to the damage spreading transition temperature 
$T_{ds}$. This is quite natural
  because the
 damage spreading is a probe for the 
ruggedness of the energy landscape.\\
Following \cite{ritort} it is interesting to note that for high values of p
 $T_{ds}$ ($\sim \sqrt{p/2}$)
 is much above the temperature $T_{TAP}$ ($\sim \sqrt{\log p}$) where 
an exponentially large number of states appears. As a consequence the origin
 of damage spreading is purely dynamical and not related to TAP states. 

\begin{figure}[bt]
\centerline{    \epsfysize=9cm
        \epsffile{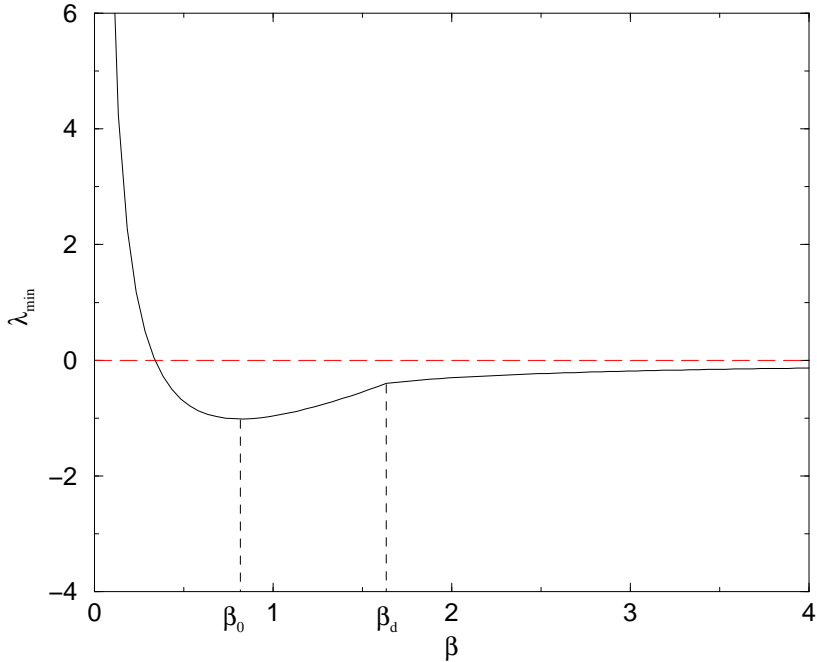}}
\caption{Left edge ($\lambda _{min}$) 
of the spectrum $\rho ^{\infty }(\mu )$ as a function of the inverse 
temperature $\beta =1/T$ for $p=3$. $1/\beta _{d}$ is the  
dynamical glass transition temperature, whereas $1/\beta _{0}$ is the 
temperature at which $\lambda _{min}$ reaches its minimum value and
 is related to the damage spreading transition.} 
\label{fig:distance}
\end{figure}

In fig.1 we have also indicated the dynamical transition temperature 
 $T_{d}=1/\beta _{d}$.
 The density of states does not change qualitatively when the temperature
 crosses $T_{d}$: a fraction of negative eigenvalue is present at $T_{d}$ and
 vanishes for T going to zero.  
Therefore there is no sign of the dynamical transition in the behaviour
 of the density of states 
of instantaneous normal modes. 

In summary, we have shown that the damage spreading transition seems to be 
related
 to a change in the local properties of the energy landscape seen by the
 system during the dynamical evolution. Conversely, the local properties 
of the energy landscape do not change qualitatively 
at the dynamical glass transition.
At the dynamical glass transition the energy landscape
 seen by the system remains locally the same, whereas its {\em global}
 properties
 change and this can be observed by analysing the {\em local} properties 
of the free energy landscape.

%%%%%%%%%%%%%%%%%%%%%%%%%%%%%%%%%%%%%%%%%%%%%%%%%%%%%%%%%%%%%%%%%%%%%%%%%%%%%%
%conclusione
%%%%%%%%%%%%%%%%%%%%%%%%%%%%%%%%%%%%%%%%%%%%%%%%%%%%%%%%%%%%%%%%%%%%%%%%%%%%%%
\section{Conclusion}\label{}

In this paper the Thouless,
Anderson and Palmer approach to thermodynamics of mean field spin
glasses has been generalised to dynamics. We have shown 
a procedure to compute dynamical TAP 
equations, which is the generalisation to dynamics of the 
$\frac{1}{d}$ ($d$ being the spatial dimension)
expansion developed by A. Georges and S. Yedidia \cite{Antoine}. This  
method has been applied to the p-spin spherical model. 
In this context we have focused
 on the interpretation of the dynamics as an evolution
 in the free energy landscape.

We have shown that at finite times the dynamics cannot be represented
 as a gradient descent in the free energy landscape, because the 
reaction term in the dynamical TAP equation is non-Markovian.\\
However, the long-time dynamics can be interpreted as an evolution
 in the free energy landscape. 
Actually, for initial conditions belonging to stable TAP states 
the long-time evolution of local magnetisations
 coincides with a gradient descent in the free energy landscape with an extra
 spherical multiplier and extra magnetic fields going to zero at 
large times. For random initial conditions the local magnetisations 
vanish asymptotically because at any finite time 
two different typical noise histories bring the system in two completely
 uncorrelated configurations \cite{Alain2}. However, also in this case,
 a description of the long-time dynamics as an evolution in the free
 energy landscape makes sense, providing that the effects due to 
the spreading of the dynamical probability measure are separated from the 
slow motion of the system.
In particular we have explicitly shown  
that slow dynamics is due to the motion in the flat directions
 of the free energy landscape in presence of a vanishing source of drift. \\
These results clarify and strengthen the relationship
 between long-times dynamics and local properties of the
 free energy landscape, 
which was already found in \cite{leticia,Franz,Alain,glauS2}.\\
Moreover we have shown that the local properties
 of the {\em energy} landscape seen during the dynamical evolution
 do not change qualitatively at the dynamical glass transition
 but at a higher temperature $T_0$, which is related to the damage spreading 
transition \cite{ritort}. This indicates that 
 at the dynamical glass transition the energy landscape
 seen by the system remains locally the same, whereas its {\em global}
 properties
 change and this can be observed by analysing the {\em local} properties 
of the free energy landscape.

Finally, we remark that there is still an 
important question which remains open and 
 which has been clarified only for the zero temperature dynamics
 of the p-spin spherical
 model \cite{laloux}: even if it is known that the TAP states 
having flat directions in the
 free energy landscape dominate the off-equilibrium dynamics, 
it is not clear why starting 
from a random initial condition the system goes toward these states.
The dynamical TAP equations are strongly non-Markovian for any finite time.
This result suggests that the matching between a certain initial
 condition (e.g. a random initial condition) and the 
asymptotic regime (slow dynamics at the threshold level) cannot
 be explained in terms of the static free energy and is a purely
 dynamical problem.
  
We conclude noticing that the formal analogies (due to the 
 superspace notation) between static and dynamic free energy
 let us hope that the study performed in this article can be extended also
 to the cases in which
 an analytical solution is not available (finite dimensional
 system), but 
 the symmetry properties of 
the asymptotic solution are known \cite{leticia2}.
 We are currently working in this direction.\\
Furthermore the dynamical TAP approach developed in this paper could
 be useful for the study of barrier crossing and instantons
 in the dynamics of mean field models \cite{ioffe}.\\
{\bf Acknowledgments}: I am deeply indebted to L. F. Cugliandolo,
 J. Kurchan and R. Monasson for
numerous, helpful and thorough discussions on this work. I wish also to thank 
 S. Franz and M. A. Virasoro 
for many interesting discussions on the asymptotic
 solution analysed in section III.B.2. I am particularly 
 grateful to R. Monasson for his constant support and for 
 a critical reading of the manuscript.

%%%%%%%%%%%%%%%%%%%%%%%%%%%%%%%%%%%%%%%%%%%%%%%%%%%%%%%%%%%%%%%%%%%%%
% Appendice 1: condizioni iniziali
%%%%%%%%%%%%%%%%%%%%%%%%%%%%%%%%%%%%%%%%%%%%%%%%%%%%%%%%%%%%%%%%%%%%%

\appendix
\section{Initial condition}\label{appendix1}
We take  as initial condition a fixed configuration $\{ s_{i}^{0} \}$;
 therefore
 in the 
Martin-Siggia-Rose functional (\ref{action0}) one has only to integrate 
on paths such that $\{ s_{i}(t=0)=s_{i}^{0}\}$. We impose this constraint 
by adding to the action (\ref{action}) the term: 
\begin{equation}\label{incon}
-\int dt \delta (t) \hat {s_{i}}(t)(s_{i}(t)-s_{i}^{0})\quad .
\end{equation}
This extra term has only two effects within the superspace
 formulation of dynamics: changes the operator $D$ and the 
(super)magnetic field $H_{i}$ in:
\begin{equation}\label{D2}
D^{in}=-2\frac{\partial ^{2}}{\partial \theta \partial \overline{\theta}}
-2\theta \frac{\partial ^{2}}{\partial \theta \partial t} + 
\frac{\partial}{\partial t}+\delta (t)\qquad , \qquad H_{i}^{in}=H_{i}-\delta (t) 
s_{i}^{0}
\end{equation}
This replacement does not affect the derivation of dynamical TAP 
equations; then taking care of the initial condition leads only to replace
 $D$ with $D^{in}$ and $H_{i}$ with $H_{i}^{in}$ in the dynamical TAP
 equations (\ref{dintapeq1} ) and (\ref{dintapeq2}). This corresponds 
to replace $\frac{\partial}{\partial t}$ with $\frac{\partial}{\partial t}
 + \delta (t) $ in equations (\ref{tc}) and (\ref{tr}) and to add to 
equation (\ref{tm}) the term $\delta (t)(m_{i}(t)-s_{i}^{0})$. \\
These new terms fix  the initial condition on magnetisations: $m_{i}(t=0)=s_{i}^{0}$ and enforce
 the equality $C(t,0)=Q(t,0)$. This last condition is already expected 
 on physical grounds because the thermal noise is not relevant at time $t=0$,
 therefore $\sum_{i=1}^{N}\left<s_{i}(t)s_{i}(0)\right>=
\sum_{i=1}^{N}\left<s_{i}(t)\right>s_{i}(0)$. \\
We note that if we do not fix any particular initial condition,
 but we take as initial condition for the dynamical measure 
an uniform average  
over all possible configurations as in \cite{leticia},
 then we find that the local magnetisations, the spherical 
multiplier, the correlation and the response functions
 fulfil 
(\ref{tc}), (\ref{tr}), (\ref{tm}) and (\ref{lambda}) without 
 the boundary condition on $C$. In this case the equations on the local 
on magnetisations are trivially satisfied because $m_{i}=0$ for every time $t$.

%%%%%%%%%%%%%%%%%%%%%%%%%%%%%%%%%%%%%%%%%%%%%%%%%%%%%%%%%%%%%%%%%%%%%%%%%%%%%
%  Appendice 2: INM 
%%%%%%%%%%%%%%%%%%%%%%%%%%%%%%%%%%%%%%%%%%%%%%%%%%%%%%%%%%%%%%%%%%%%%%%%%%%%%
\section{INM}\label{appendix2}
In this Appendix we show a standard way to compute the density of
 states of ${\mathbf{E}}$:
\begin{equation}\label{e}
E_{i,j}=\frac{p (p-1)}{p!}\sum_{i_{3}\neq \dots \neq i_{p}}J_{i,j,i_{3},\dots ,i_{p}}
s_{i_{3}}\cdots s_{i_{p}}\quad ,
\end{equation} 
where $\{ s_{i}(t)\}$ is an instantaneous dynamical configuration.\\
The spectral properties of ${\mathbf{E}}$ can be obtained through the
knowledge of the resolvent $G (\mu + i \epsilon)$, that is the trace of
$((\mu + i\epsilon ) {\mathbf{1}}-{\mathbf{E}})^{-1} $ \cite{report}. 
Denoting the average over disorder by $\overline{(\cdot )}$ and the
 average over instantaneous configurations at time $t$ by 
$\left< \cdot \right>$, the mean
density of states reads 
\begin{equation}
\rho  (\mu ;t )= - \frac{1}{\pi} \lim _{\epsilon \to 0 ^+} 
\hbox{\rm Im} \;\overline{ \left<G (\mu  + i \epsilon )\right> } \quad .
\label{spectre}
\end{equation}
The averaged resolvent is then written as the propagator of a
replicated Gaussian field theory \cite{report}
\begin{eqnarray}
\overline{G (\mu + i \epsilon ) } &=& \lim_{n\rightarrow 0}
\frac{-i}{Nn}  \int \prod _{i} d\vec \phi _{i} 
\sum_{k=1} ^N\vec \phi _k ^{\; 2} 
\overline{\left<\exp (L(s_{i}(t),\mu))\right>}  
\nonumber \\ \hbox{\rm where} \quad
L(s_{i}(t),\mu)&=&\sum_{i=1}^{N}\frac{i}{2}( \mu + i\epsilon )
\vec \phi _{i} ^{\;2} -\sum_{i,j=1}^{N}\frac{i}{2} E_{i,j}
\vec {\phi}_{i}\cdot\vec {\phi}_{j} \quad . 
\label{field}
\end{eqnarray}
Replicated fields $\vec \phi _i$ are $n$-dimensional vector fields
attached to each site $i$. The average over the instantaneous dynamical
configurations can be written in terms of Martin-Siggia-Rose functional 
\cite{msr}: 
\begin{eqnarray}\label{LMSR}
\left<\exp (L(s_{i}(t),\mu))\right>&=&\int\prod_{i=1}^{N}{\cal D}s_{i}
{\cal  D}\hat{s}_{i}
\exp \left[ L(s_{i},\mu)+S_{MSR}(s_{i},\hat{s}_{i}) \right] \\
S_{MSR}(s_{i},\hat{s}_{i})&=&\int_{0}^{+\infty} dt \sum_{i=1}^{N}-\hat{s}_{i}\left(\frac{ds_{i}}{dt}+\frac{\partial H}
{\partial s_{i}}-T\hat{s_{i}}\right)\quad ,
\end{eqnarray}
where we do not write the ``ghosts'' fields because we take the Ito
 convention.\\
Within the Martin-Siggia-Rose approach the average over the couplings is
 a simple Gaussian integral. Then we find that the total action is
 equal to:
\begin{eqnarray}\label{dopomedia}
&&S(s_{i},\hat {s}_{i},\phi _{i}^{a},t)=\sum_{i=1}^{N}\frac{i}{2}( \mu + i\epsilon )\vec \phi _{i} ^{\;2}+ 
\int_{0}^{+\infty}dt \sum_{i=1}^{N}-\hat{s}_{i}\left(\frac{ds_{i}}{dt}
-T\hat{s_{i}}+\lambda s_{i}\right)+\nonumber \\
&&+\frac{1}{4 N^{p-1}}\sum_{i_{1}\neq \cdots \neq  i_{p}}\left[ 
\int dt'\left(\hat{s}_{i_{1}}s_{i_{2}}\cdots s_{i_{p}}+
\cdots +s_{i_{1}}\cdots s_{i_{p-1}}\hat{s}_{i_{p}}+\frac{i}{2}
\sum_{l\neq m}O_{l,m}^{t}  \right) \right]^{2}\quad ,
\end{eqnarray}
where $O_{l,m}^{t}$ is symmetric under the exchange of $l$ and $m$ and
 is equal to:
\begin{eqnarray}\label{Olm}
O_{l,m}^{t}(t')&=&\delta (t'-t)\sum_{a=1}^{n}s_{i_{1}}(t')\cdots 
 s_{i_{l-1}}(t')\phi _{i_{l}}^{a}
s_{i_{l+1}}(t')\cdots s_{i_{m-1}}(t')\phi _{i_{m}}^{a}s_{i_{m+1}}(t')\cdots s_{i_{p}}(t') 
\qquad (l<m)\\
O_{l,m}^{t}(t')&=&0 \qquad l=m\quad .
\end{eqnarray}
The action (\ref{dopomedia}) depends on $s_{i},\hat {s}_{i},\phi _{i}^{a}$
 only through
\begin{eqnarray}\label{parametri}
C(t,t')=\frac{1}{N}\sum_{i=1}^{N}s_{i}(t)s_{i}(t') \quad 
&R(t,t')=\frac{1}{N}\sum_{i=1}^{N}s_{i}(t)\hat {s}_{i}(t')&\quad 
D(t,t')=\frac{1}{N}\sum_{i=1}^{N}\hat{s}_{i}(t)\hat {s}_{i}(t')\\
K_{a}(t)=\frac{1}{N}\sum_{i=1}^{N}s_{i}(t)\phi _{i}^{a}\quad 
&\hat {K}_{a}(t)=\frac{1}{N}\sum_{i=1}^{N}\hat {s}_{i}(t)\phi _{i}^{a}&\quad
Q_{a,b}=\frac{1}{N}\sum_{i=1}^{N}\phi _{i}^{a}\phi _{i}^{b}\quad .
\end{eqnarray}
In the large-N limit the functional integral giving the resolvent
 $G$ is dominated by a saddle point contribution.\\
The action (\ref{dopomedia}) is invariant when one changes 
$\vec{\phi }\rightarrow -\vec{\phi }$, therefore we take 
$K_{a}(t)=0$ and $\hat {K}_{a}(t)=0$. With this choice the 
 saddle point equations on $K_{a}(t)$ and $\hat {K}_{a}(t)$ are 
trivially satisfied.\\
The saddle point equations on $C(t,t')$, $R(t,t')$ and $D(t,t')$ are 
 the usual ones \cite{leticia}.\\
Using the identity
\begin{equation}\label{relaz}
-\delta _{a,b}=\frac{1}{N}\sum_{i=1}^{N}
<<\phi _{i}^{a}\frac{\partial S}{\partial \phi _{i}^{b}}>>\quad ,
\end{equation}
where $<<\cdot >>$ means the average over $s_{i},\hat {s}_{i},\phi _{i}^{a}$
(with weight $\exp (S)$), it is easy to obtain 
the saddle point equation on $Q_{a,b}$:
\begin{equation}\label{qab}
\frac{1}{2} p (p-1)C(t,t)^{p-2}\sum_{c=1}^{n}Q_{b,c}Q_{a,c}
-i(\mu  +i\epsilon )Q_{a,b}-\delta _{a,b}=0\quad .
\end{equation}
It is well known that in the computation of the density of states
the replica symmetric saddle point gives the leading contribution. 
Therefore we take $Q_{a,b}=q\delta _{a,b}$. \\
The spherical constraint fixes $C(t,t)=1$, then from 
the saddle point equation 
 on $q$ we get $q=\frac{i\mu+\sqrt{2p (p-1)-\mu  ^{2}}}{p (p-1)}$.\\
The density of states of $\mathbf{E}$ is obtained from (\ref{spectre}): 
\begin{equation}\label{semicerchio}
\rho  (\mu;t)=\frac{1}{\pi}\Re (q)=\frac{1}{\pi p(p-1)}
\sqrt{2 p (p-1)-\mu ^{2}}
\end{equation}
which is the Wigner semi-circle law. As expected this density does not
 depend on $t$. The only dependence on time for the density
 of states of the energy Hessian comes from $\frac{\lambda (t)}{\beta }$, which
 translates the centre of the semi-circle (see eq. (\ref{semicirc})).\\

%%%%%%%%%%%%%%%%%%%%%%%%%%%%%%%%%%%%%%%%%%%%%%%%%%%%%%%%%%%%%%%%%%%%%%%%%%%%%%%%%%%%%%%%%%%%%%%%%%%%%%%%%%%%%%%%%%%%%%%%%%%%%%%%%%%%%%%%%%%%%%%%%%%%%%%%%%%%%%


\begin{thebibliography}{99}

\bibitem{rivista}
For a recent review, see~:\\
J.P. Bouchaud, L. Cugliandolo, M. M{\'e}zard, J. Kurchan, {\em ``Out of 
equilibrium dynamics in spin-glasses and other glassy systems''}, 
A.P. Young editor, World Scientific, Singapore (1997) 
and references therein.

\bibitem{landscape}
See for instance many contributions in: {\em 
``Landscape Paradigms in Physics and Biology'', 
Physica D {\bf 107}, 117-437 (1997)}.

\bibitem{beyond}
M. M{\'e}zard, G. Parisi and M. Virasoro, {\em 
Spin Glass Theory and Beyond (1987)} (Singapore: World Scientific).

\bibitem{TAP}
D. J. Thouless, P. W. Anderson and R. G. Palmer, {\em Phil. Magazine {\bf 35}, 593 (1977)}.

\bibitem{SK}
D. Sherrington and S. Kirkpatrick, {\em Phys. Rev. Lett. {\bf 35}, 1972 (1975)}.

\bibitem{young}
C. De Dominicis and A.P. Young, {\em J. Phys. A {\bf 16}, 2063 (1983)}.

\bibitem{cavita}
M. M{\'e}zard, G. Parisi and M. Virasoro, {\em Europhys. Lett. {\bf 1}, 77 (1986)}.

\bibitem{Parisi}
G. Parisi, {\em Phys. Lett. A {\bf 73}, 154 (1979)}, 
 {\em J. Phys. A {\bf 13}, L115 (1980)}, 
{\em ibid {\bf 13}, 1101 (1980)}, {\em ibid {\bf 13}, 1887 (1980)}.

\bibitem{leticia}
L.F. Cugliandolo and J. Kurchan, {\em Phys. Rev. Lett. {\bf 71}, 173 (1993)}
 and {\em Phil. Mag. B {\bf 71}, 501 (1995)}.

\bibitem{Remi}
R. Monasson, {\em Phys. Rev. Lett. {\bf 75}, 2847 (1995)}.

\bibitem{Franz}
S. Franz and G. Parisi, {\em J. Phys. (France) I {\bf 5}, 1401 (1995)}.

\bibitem{laloux}
J. Kurchan and L. Laloux {\em J. Phys. A {\bf 29}, 1929 (1996)}.

\bibitem{vetri1}
T. R. Kirkpatrick and P. G. Wolynes, {\em Phys. Rev. A {\bf 34}, 1045 (1986)};
 T. R. Kirkpatrick and D. Thirumalai, {\em Phys. Rev. Lett {\bf 58}, 2091 (1987)};T. R. Kirkpatrick and D. Thirumalai, {\em Phys. Rev. B {\bf 36}, 5388 (1987)}; T. R. Kirkpatrick and D. Thirumalai and P. G. Wolynes, {\em Phys. Rev. A {\bf 40}, 1045 (1989)}.

\bibitem{vetri2}
G. Parisi and M. M{\'e}zard, {\em Phys. Rev. Lett. {\bf 82},
747 (1999)}.

\bibitem{mcep-spin}
 J. P. Bouchaud, L. F. Cugliandolo, J. Kurchan and M. M{\'e}zard, {\em Physica A {\bf 226}, 243 (1996)}.

\bibitem{gotze}
W. G{\"o}tze, {\em J. Phys. Condens. Matter {\bf 11}, A1-A45 (1999)}.

\bibitem{ptap}
J. Kurchan, G. Parisi and M. A. Virasoro, {\em J. Phys I (France) {\bf 3}, 1819 (1993)}.

\bibitem{crisantitap}
A. Crisanti and H.-J. Sommers, {\em J. Phys. I France {\bf 5}, 805 (1995)}.

\bibitem{crisantidyn}
A. Crisanti, H. Horner and  H.-J. Sommers, 
{\em Z. Phys. B {\bf 92}, 257 (1993)}.

\bibitem{Alain}
A. Barrat, R. Burioni and M. M{\'e}zard , {\em J. Phys. A {\bf 29}, L81 (1996)}.

\bibitem{keyes}
T. Keyes {\em J. Phys. Chem. A {\bf 101}, 2921 (1997)}.

\bibitem{ritort}
M. Heerema and F. Ritort {\em J. Phys. A: Math. Gen. {\bf 31}, 8423
(1998)} and cond-mat/9812346.

\bibitem{tapstatiche}
H. J. Sommers, {\em Z. Phys. B {\bf 31}, 301 (1978)}; C. De Dominicis,
{\em Phys. Rep. B {\bf 67}, 37 (1980)}; H. Rieger, {\em Phys. Rev. B {\bf 46}, 14655 (1992)} .

\bibitem{doubletransform}
J. M. Cornwall, R. Jackiw and E. Tomboulis, {\em Phys. Rev. {\bf 10}, 2428 
(1974)}; R. W. Haymaker, {\em Riv. Nuovo Cimento {\bf 14}, 1 
(1991)}.

\bibitem{Plefka}
T. Plefka, {\em J. Phys. A {\bf 15}, 1971 (1982)}.

\bibitem{Antoine}
A. Georges and J.S. Yedidia, {\em J. Phys. A {\bf 24}, 2173 (1991)}.

\bibitem{rem}
D.J. Gross and M. M{\'e}zard, {\em Nuc. Phys. B {\bf 240},
431 (1984)}. 

\bibitem{Alain3}
A. Barrat  {\em cond-mat/9701031, unpublished}.

\bibitem{crisantirep}
A. Crisanti and  H.-J. Sommers, {\em Z. Phys. B {\bf 87}, 341 (1992)}.

\bibitem{kurfranz}
J. Kurchan, {\em J. Phys. France {\bf 2}, 1333 (1992)};
S. Franz and J. Kurchan, {\em Europhys. Lett. {\bf 20}, 197 (1992)}.

\bibitem{zj}
J. Zinn-Justin, {\em Quantum Field Theory and Critical Phenomena 
}, Clarendon Press 1997.

\bibitem{msr}
P.C. Martin, E.D. Siggia and H.A. Rose, {\em Phys. Rev. A {\bf 8}, 423 (1978)};
C. de Dominicis and L. Peliti, {\em Phys. Rev. B {\bf 18}, 353 (1978)}.

\bibitem{gozzi}
E. Gozzi, {\em Phys. Lett. {\bf 143}, 183 (1984)}.

\bibitem{somzip}
H. Sompolinsky and A. Zippelius, {\em Phys. Rev. B {\bf 25}, 6860 (1982)}.

\bibitem{ioffe}
L. B. Ioffe and D. Sherrington, {\em Phys. Rev. B {\bf 57}, 7666 (1998)}.
A. V. Lopatin and L. B. Ioffe, {\em Phys. Rev. B {\bf 60}, 6412 (1999); cond-mat/9907135}.

\bibitem{Bouchaud}
J. P. Bouchaud, {\em J. Phys. I (France) {\bf 2} 1705 (1992)}.
\bibitem{Alain2}
A. Barrat, R. Burioni and M. M{\'e}zard, {\em J. Phys. A {\bf 29}, 1311 (1996)}.

\bibitem{qe}
S. Franz and M. A. Virasoro, {\em condmat/9907438}.

\bibitem{preparation}
G. Biroli, S. Franz and M. A. Virasoro, {\em in preparation}.

\bibitem{Dean}
L. F. Cugliandolo and D. S. Dean, {\em J. Phys. A {\bf 28}, 4213 (1995)}.
 
\bibitem{moore}
A. Bray and M. A. Moore, {\em J. Phys. C: Solid State Phys. {\bf 12}, L441 
(1979)}.

\bibitem{mehta}
M. L. Metha, {\em Random Matrices and the Statistical theory of Energy Levels 
(1967)} (New York: Academic).

\bibitem{glauS2}
G. Biroli and R. Monasson, {\em J. Phys. A {\bf 31}, L391 (1998)}.

\bibitem{leticia2}
L.F. Cugliandolo and J. Kurchan, {\em Physica A {\bf 263}, 242 (1999)}.

\bibitem{report}
D.J. Thouless, {\em Physics Reports {\bf 13}, 93 (1974)}; Mesoscopic 
quantum physics, Les Houches session LXI, (Elsevier Science, 1995).

\end{thebibliography}
\end{document}